\documentclass[preprint]{aastex}

\let\captionbox=\undefined

\usepackage{graphicx}
\usepackage{caption}
\usepackage{subfigure}
\usepackage{multirow}
\usepackage{amsmath}

\begin{document}

\title{
 {\sl Suzaku} Observations of the X-ray Brightest Fossil Group ESO 3060170 
}

\author{Yuanyuan Su\altaffilmark{1}, Raymond E. White III\altaffilmark{1}, and Eric D. Miller\altaffilmark{2}} 

\altaffiltext{1}{Department of Physics and Astronomy, University of 
Alabama, Box 870324, Tuscaloosa, AL 35487, USA}

\altaffiltext{2}{Kavli Institute for Astrophysics \& Space Research, Massachusetts Institute of Technology, Cambridge, MA 02139, USA}

\email{ysu@crimson.ua.edu}



\begin{abstract}
``Fossil" galaxy groups, each dominated by a relatively isolated giant elliptical galaxy, have many properties intermediate between groups and clusters of galaxies. We used the {\sl Suzaku} X-ray observatory to observe the X-ray brightest fossil group, ESO~3060170, out to $R_{200}$, in order to better elucidate the relation between fossil groups, normal groups, and clusters. We determined the intragroup gas temperature, density, and metal abundance distributions and derived the entropy, pressure and mass profiles for this group. The entropy and pressure profiles in the outer regions are flatter than in simulated clusters, similar to what is seen in observations of massive clusters. This may indicate that the gas is clumpy and/or the gas has been redistributed. Assuming hydrostatic equilibrium, the total mass is estimated to be $\sim1.7\times10^{14}$ $M_{\odot}$ within a radius $R_{200}$ of $\sim1.15$ Mpc, with an enclosed baryon mass fraction of 0.14.
The integrated iron mass-to-light ratio of this fossil group is larger than in most groups and comparable to those of clusters, indicating that this fossil group has retained the bulk of its metals. A galaxy luminosity density map on a scale of 25 Mpc shows that this fossil group resides in a relatively isolated environment, unlike the filamentary structures in which typical groups and clusters are embedded.
\end{abstract}


\section{\bf Introduction}
\smallskip

In the hierarchical structure formation scenario, galaxy systems evolve through mergers 
and accretion, with galaxy groups regarded as the building blocks of galaxy clusters.
However, X-ray observations of groups and clusters suggest that 
clusters are not simply scaled-up versions of present-day groups.
To date, most X-ray observations of the gas in groups and clusters of galaxies 
have been restricted to within $\sim R_{500}$, the radius (typically $0.4-1.3$ Mpc)
within which the average density is 500 times the cosmological critical density.  
Based on X-ray observations within $\sim R_{500}$,
clusters appear to differ from groups in several ways.
1) The baryonic mass fraction in clusters varies only weakly with cluster temperature, 
with values nearly equal to the cosmological value of $f_b=0.15$ ({\sl Planck} Collaboration 2013);
however, the baryon fraction in groups ranges widely ($f_b\sim0.04-0.1$), 
with values typically much smaller than the cosmological value and rising with
group temperature (Dai et al.\ 2010).
2)  Clusters have about the same ratio of gaseous iron mass to stellar light, 
regardless of their temperature, while
in groups the iron-mass-to-light ratio (IMLR) drops precipitously with gas temperature 
when $kT\lesssim 1$ keV;  galaxy groups have up to 50 times less iron per unit optical luminosity than clusters (Renzini 1997).
3) Scaling relations involving X-ray luminosity, gas
temperature, entropy, and mass differ between clusters and groups (Voit et al.\ 2005).
Based on the studies cited above, we adopt $\sim$ 2 keV for the boundary between
galaxy groups and clusters.

All of these differences between groups and clusters may have the same origin:
the gravitational potential wells of clusters are deep enough that they have 
retained the bulk of their baryons, including gas enriched by supernovae ejecta;
however, the shallower potential wells of groups make them more vulnerable to 
gas loss due to energy input from nongravitational processes such as
galactic winds and feedback from active galactic nuclei (AGN) (Sun 2012). 
This account needs to be confirmed with deep observations of the outer 
parts of galaxy groups to test whether 
the ``missing" baryons and metals can be found in the outer atmospheres of groups.

Thanks to a low and stable instrumental background, the
{\sl Suzaku} X-ray satellite observatory has been able to 
observe a number of massive clusters out to
their virial radii or even beyond  (Bautz et al.\ 2009; George et al.\ 2009; 
Hoshino et al.\ 2010; Kawaharada et al.\ 2010; Simionescu et al.\ 2011; Reiprich et al.\ 2009; Sato et al.\ 2012; Walker et al.\ 2012a,b).
Baryon mass fractions are observed to reach the cosmological value 
or even greater (Simionescu et al.\ 2011; Sato et al.\ 2012).  
Metals are reported to be found at the virial radii of several clusters
(Fujita et al.\ 2008; George et al. 2009; Simionescu et al. 2011; Urban et al.\ 2011). 
Surprisingly, gaseous entropy profiles are found to flatten beyond 
$\sim$ $R_{500}$, contrary to expectations from numerical simulations which show 
entropy rising nearly linearly with radius: $S\sim r^{1.1}$ (Voit et al.\ 2005).
The observed entropy flattening has been interpreted in various ways:
1) it may be due to clumped gas, 
which would increase the emission-measure-weighted gas density 
over the actual average gas density (Simionescu et al.\ 2011);
2) the entropy flattening may arise
from electrons being out of thermal equilibrium with ions, due to recent accretion
shocks in cluster outskirts (Hoshino et al.\ 2010; Akamatsu et al.\ 2011); 
3) the weakening of accretion shocks in relaxed clusters may reduce the entropy at large radii 
(Cavaliere et al.\ 2011, Walker et al.\ 2012c).
Despite there being many more groups than clusters, there are currently
not many X-ray observations of groups or poor clusters out to their virial radii, due to 
their relatively low surface brightness.

A particularly interesting subset of such galaxy systems are fossil groups, 
which are empirically defined as systems with 1) a central
dominant galaxy more than two optical magnitudes brighter than the second
brightest galaxy within half a virial radius; and 2) an extended thermal
X-ray halo with $L_{\rm (X,bol)}$ $>$ $10^{42}$ $h_{50}^{-2}$ erg s$^{-1}$ (Jones et al. 2003).
A fossil group is thought to be the evolutionary end-result of a group of galaxies having
mostly merged into one giant elliptical (Mulchaey \& Zabludoff 1999). 
Fossil groups are not rare ---
they are estimated to constitute 8-20\% of groups and clusters (Jones et al.\ 2003). 
Yang et al.\ (2008) used the Sloan Digital Sky Survey (SDSS) 
to show that 20\%--60\% of groups with 
masses $\approx10^{13}M_{\odot}$ are fossil groups. 
Simulations predict that about 33\% of groups are fossil groups (D'Onghia et al.\ 2005).
The properties of fossil groups generally populate the transition regions
in the scaling relations for groups and clusters
(Khosroshahi et al.\ 2007; Proctor et al.\ 2011; Miller et al.\ 2012).
More than half  have $kT\gtrsim2$ keV, 
which is high for groups and more typical of poor clusters
(Khosroshahi et al.\ 2007; Vikhlinin et al.\ 1999).

There are multiple lines of evidence suggesting that fossil groups tend to
be old, relaxed systems.
For example,
X-ray observations show that fossil groups are associated with unusually concentrated 
dark matter distributions (Khosroshahi et al.\ 2007; von Benda-Beckmann et al.\ 2008), which are associated 
with earlier formation epochs in numerical simulations (Giocoli et al.\ 2012;  Dariush et al.\ 2007).
Secondly,
the optical isophotes and X-ray atmospheres of fossil groups tend to appear undisturbed, 
implying that their last significant mergers were long ago. 
Moreover, simulations by 
D'Onghia et al.\ (2005) showed that groups with earlier formation times
have larger luminosity differences between their 1st and 2nd brightest galaxies.
However, some observational evidence suggests that fossil groups may have formed 
or been significantly disturbed more recently.
For example, fossil groups tend not to have very extensive cool cores, despite their central
cooling times being less than the Hubble time (Miller et al.\ 2012).
In addition, Dupke et al.\ (2010) showed that there is an enhancement of SNII ejecta 
in the central regions of some fossil groups, perhaps due to star formation triggered
by recent mergers forming their central dominant galaxies. 
Interestingly, von Benda-Beckmann et al.\ (2008) showed in simulations tracking fossil groups 
forward in time that the majority had their luminosity gap (between 1st and 2nd-brightest galaxies) 
filled by new infalling galaxies; 
this suggests that fossil groups are a transitional phase, not the stable end-result of a merging group. 

To understand the evolution and nature of fossil groups, it is
crucial to understand how their structure and environment compare to those
of non-fossil systems.  A key strategy is to explore the outer regions of
the group gas, using similar methods to those described
above.  This has been previously done for  fossil group RXJ 1159+5531
by Humphrey et al.\ (2012), who trace the group gas out to $R_{\rm vir} = R_{108}$.
They find no flattening of the entropy profile and no need to invoke
gas clumping, indicating that this phenomenon might depend on the
environment or evolutionary state of the system. 
The baryon fraction of RXJ 1159+5531 is consistent with the cosmological value, much larger than what has been observed in normal groups at smaller radii.
Studies of additional systems are clearly needed to further constrain the
fossil group paradigm. 

To learn more about the nature of fossil groups,
we performed a deep study of the X-ray brightest fossil group, ESO~3060170,
with {\sl Suzaku}.  
ESO 3060170 was first selected as a fossil group candidate from a sample of early-type galaxies observed with
the {\sl ROSAT} X-ray observatory (Beuing et al.\ 1999). Its second brightest galaxy within half a virial radius is 2.5 magnitudes fainter in the $R$-band than the central dominant galaxy, qualifying it as a fossil group (Sun et al.\ 2004).
ESO~3060170  was subsequently observed with {\sl Chandra} and {\sl XMM-Newton} out to 450 kpc 
($\approx 0.33$ $R_{\rm vir}$) by Sun et al.\ (2004), who found it to be the most massive fossil group known. 
It is also the X-ray brightest known fossil group{\footnote{Some consider NGC 1550 to be the X-ray brightest fossil group, 
but Sun et al.\ (2003) showed that it has 3 galaxies  within 0.5 $R_{\rm vir}$  that are within 2 $R$ magnitudes of NGC 1550, so this group does not meet the usual criterion for a fossil group.}.
It has an X-ray luminosity of 
$2.5\times10^{43}$ ergs s$^{-1}$  (0.5-2.0 keV) within 0.33  $R_{\rm vir}$  (Sun et al.\ 2004),
making it an ideal candidate for detection of group gas out to its virial radius.  
Adopting a redshift of $z=0.0358$ from the NASA/IPAC Extragalactic Database (NED),
we derive a luminosity distance of 153 Mpc (so $1^{\prime} = 41$ kpc), 
assuming a  cosmology with $H_0=70$ km s$^{-1}$ Mpc$^{-1}$,
$\Omega_{\Lambda}=0.7$ and  $\Omega_m=0.3$.
We studied this group out to $R_{200}=27.5^\prime$ with {\sl Suzaku}, in order to determine
its gaseous temperature, abundance, pressure and density profiles at large radii,
as well as estimate its total mass and baryon fraction.  
From the total mass density distribution described in $\S3.6$, we determined
that $R_{500}=750$ kpc and  $R_{200}=1.15$ Mpc.
We describe the observations and data reduction techniques in \S2 and report results regarding the thermal state and metal content in \S3; 
their implications are discussed in \S4 and our main conclusions are summarized in \S5.

\medskip
    
\section{\bf Data reduction} 
\smallskip

We  observed the central and outer regions of ESO~3060170 with {\sl Suzaku}
in two pointings during May 2010 (PI: Y.~Su).
Three X-ray CCDs (XIS 0, 1 and 3) were used and
the observation log is listed in Table~1. 
A mosaic of the XIS images in the 0.5-4.0 keV band is shown in Figure~\ref{fig:eso}. 
This observation was taken after XIS0 was partially damaged
by being  flooded with large amounts of charge, which saturated the electronics  
and created a permanently unusable 
region\footnote{http://www.astro.isas.jaxa.jp/suzaku/doc/suzakumemo/suzakumemo-2010-01.pdf}. 

Data reduction and analysis were performed with the {\sl HEASOFT} 6.12 software package 
using CalDB 20120209. 
Data were obtained in both $3\times3$ and $5\times5$ data readout modes;
$5\times5$  mode data were converted to $3\times3$ mode and merged with the $3\times3$ mode data. 
The events were filtered by retaining those with a geomagnetic cutoff rigidity $>6$ GeV/c, 
and an Earth elevation $>10^\circ$.
The calibration source regions and hot pixels were excluded. 
Light curves were filtered using {\sl CIAO} 4.3 script {\tt lc\_clean} to exclude times with flares. 
The resulting effective exposure times are shown in Table~1. 
During these observations, the activity of the Solar Wind Charge eXchange (SWCX) 
was inferred to be low, based on data from the {\sl Advanced Composition Explorer} (ACE).  
Stray light should be negligible since this target is not extremely bright 
and the two pointings were taken with a diagonal offset, which minimizes the
effects of stray light (Takei et al.\ 2012).

\subsection{\sl Image Analysis} 

To produce a vignetting-corrected image, we followed standard 
procedure\footnote{http://heasarc.gsfc.nasa.gov/docs/suzaku/analysis/expomap.html}
by generating flat fields based on ray-tracing
the Cosmic X-ray Background (CXB) spectrum through the telescope optics. 
We first fit a spectral model to the CXB in a background region described in $\S2.2$. 
Based on this spectrum, we used the FTOOL {\tt mkphlist} to 
simulate $10^{8}$ photons between $0.5-4.0$ keV, representing a uniform, extended source with radius $20^\prime$. 
These simulated incident photons were used as the input to an
X-ray telescope ray-tracing simulator (FTOOL {\tt xissim}; Ishisaki et al.\ 2007) 
to produce a flat-field image for each chip and pointing. 
We used the {\sl CIAO 4.3} tool {\tt csmooth} to smooth these flat fields 
with a Gaussian kernel (of 5 pixels) 
in order to increase the signal-to-noise ratio. 
We scaled these vignetting-corrected images by the relevant exposure times to 
produce exposure maps.  
Source images from each chip and pointing were extracted in the 0.5-4.0 keV band. 
A particle (non-X-ray) background (NXB) component was generated for each chip 
and pointing with FTOOL {\tt xisnxbgen} (Tawa et al.\ 2008). 
Mosaics of the source images, NXB images, and exposure maps for each chip and pointing were 
cast on a common sky coordinate reference frame. 
The NXB component was subtracted from each cluster image on each chip;
these images were then divided by their exposure maps and merged. Finally, we subtracted the CXB surface brightness from the image. The exposure-corrected, NXB- and CXB-subtracted 0.5-4.0 keV image mosaic is shown in Figure~\ref{fig:eso}.
Two point sources in the offset pointing, identified by eye and indicated in Figure~\ref{fig:eso}, 
were excluded from the spectral analysis.

\subsection {\sl Spectral analysis } 

We extracted spectra from six truncated annular regions, indicated in Figure~\ref{fig:eso},
centered on ESO 3060170
and extending south from
$0^\prime$-$3^\prime$, $3^\prime$-$8^\prime$, $8^\prime$-$14^\prime$, 
$14^\prime$-$20^\prime$, $20^\prime$-$25^\prime$ and $25^\prime$-$30^\prime$ (regions 1--6).  
The FTOOL {\tt xissimarfgen} was used to generate an ancillary response file 
(ARF) for each region and detector. To provide the appropriate photon weighting
for each ARF, we used a  
2-dimensional $\beta$-model surface brightness distribution extrapolated from {\sl XMM-Newton} observations (Sun et al.\ 2004). We used the C-statistic in all our spectral analyses 
and accordingly grouped  spectra to have at 
least one count per energy bin.

Redistribution matrix files 
(RMFs) were generated for each region and detector using the FTOOL {\tt xisrmfgen}. 
NXB spectra were generated with the FTOOL {\tt xisnxbgen}. 
Spectra from XIS0, 1, \& 3 were simultaneously fit  in spectral modeling 
with {\sl XSPEC} v12.7.2 (Arnaud 1996). 
We adopted the solar abundance standard of Asplund at al.\ (2009) in thermal spectral models. 
To compare our results with those using the abundance 
standard of Anders \& Grevesse (1989), our iron abundances should be multiplied by 0.67.
Energy bands were restricted to 0.5-7.0 keV for the back-illuminated CCD (XIS1) and 0.6-7.0 keV for the front-illuminated CCDs (XIS0, XIS3), 
where the responses are best calibrated. 
Uncertainties in fitted parameters are reported at the 90\% confidence level.  

\subsubsection{X-ray background}
We used three different methods to determine the X-ray background. 
In the first method, called ``offset" background,
we use background spectra (one for each detector) extracted
from a region $\sim28^\prime$ south of the center (partially overlapping region 6).
Since the observed surface brightness profile is flat in this background region, 
ARFs were generated  by assuming uniform sky emission with a radius of $20^\prime$.
We simultaneously fitted these background spectra to determine 
the surface brightness of Galactic components and the CXB, taking into account
that there may be small amounts of emission from the group at this distance. 
We used {\sl XSPEC} to fit each spectrum with a multi-component background model 
consisting of an {\tt apec} thermal emission model for the Local Bubble 
({\tt apec$_{\rm LB}$}), 
an additional {\tt apec} thermal emission model ({\tt apec$_{\rm MW}$}) for the Milky Way emission  
in the line of sight  (Smith et al.\ 2001), 
and a power law model {\tt pow$_{\rm CXB}$} (with index $\Gamma=1.41$) characterizing the CXB (De Luca \& Molendi 2004).  We also included a 
thermal {\tt apec} model for any residual emission from the group 
({\tt apec$_{\rm ESO}$}) in this outer region.
All these components but the Local Bubble were assumed to be absorbed
by foreground (Galactic) cooler gas, with the absorption characterized by 
the {\tt phabs} model for photoelectric absorption.  
Photoionization cross-sections were from Balucinska-Church \& McCammon (1998). 
We adopted a Galactic 
hydrogen column of $N_H=3.15\times10^{20}$ cm$^{-2}$ toward ESO 3060170, deduced 
from the Dickey and Lockman (1990) map incorporated in the {\sl HEASARC}
$N_H$ tool.
The final spectral fitting model was 
$ {\tt apec}_{\rm LB}+{\tt phabs} \times ({\tt apec}_{\rm MW}+{\tt pow_{\rm CXB}}+{\tt apec}_{\rm ESO}). $
The temperatures of the Milky Way and Local Bubble
components were fixed at 0.25 keV and 0.10 keV, 
respectively (Kuntz \& Snowden 2000).

The second method we call ``local" background.  
We fitted the spectra of regions 2 through 6 simultaneously  with the source plus background model 
${\tt apec}_{\rm LB}+{\tt phabs}\times({\tt apec}_{\rm MW}+{\tt pow_{\rm CXB}}+{\tt apec}_{\rm ESO})$.  
We did not include region 1 because the contribution from low mass X-ray binaries (LMXBs) in the central region may be degenerate 
with the CXB component. 
X-ray background components ({\tt apec}$_{\rm LB}$, {\tt pow$_{\rm CXB}$}, {\tt apec}$_{\rm MW}$) were forced to have the same surface brightness in all regions. 
Group gaseous components were allowed to vary independently for each region. 

The third method we label ``deproj" background, which we determined 
through deprojected spectral fits. 
We performed a deprojected spectral analysis for regions 1--6 simultaneously 
using models described later in \S2.2.3.   
Similar to the second method, we let the normalizations of all components, 
including  group emission and X-ray background components (CXB, MW and LB), 
free to vary. 
X-ray background components ({\tt apec}$_{\rm LB}$, {\tt pow$_{\rm CXB}$}, 
{\tt apec}$_{\rm MW}$) were forced to 
have the same surface brightness in all regions, while
group emission was allowed to vary independently in each region.

The results for the background components determined with these three methods are shown in Table~2. 
They are largely consistent with each other and are in line with the values 
determined by Yoshino et al.\ (2009) for 14 fields observed with {\sl Suzaku}.  
Since our main results are obtained with deprojected spectral fits,
we adopted the third method (``deproj") to determine the X-ray backgrounds.
We performed Markov chain Monte Carlo (MCMC) simulations with 40 chains,  each chain having 10000 steps, 
for the simultaneous fitting used in the third method. 
With the {\tt margin} command in {\it XSPEC}, we marginalized over the 
uncertainties of the normalizations of the X-ray background components. 
We obtained the probability distribution for each background component normalization
to determine its mean and error (listed in the starred lines of Table 2).

\subsubsection{Projected spectral analysis}
We fitted the spectrum of regions 2-6 with the following model: 
${\tt apec}_{\rm LB}+{\tt phabs}\times({\tt apec}_{\rm MW}+{\tt pow_{\rm CXB}}+{\tt apec}_{\rm ESO}).$
Here the {\tt apec}$_{\rm LB}$, {\tt apec}$_{\rm MW}$, and {\tt pow$_{\rm CXB}$} models
represent background components discussed earlier, while
the {\tt apec}$_{\rm ESO}$ component represents thermal emission from group gas.
For the central region 1 only, we added a power law component
({\tt pow$_{\rm LMXB}$}) with an index of 1.6 (and variable normalization) to 
represent unresolved LMXBs
in the central galaxy (Irwin et al.\ 2003).  ARFs derived from the $\beta$-model image  
were used for each region. 
Each background component was fixed at the surface brightness 
determined earlier. XIS1 spectra of regions 1--6 are shown in Figure~\ref{fig:spectra}. 
Its associated fitting results are listed in Table~3 as indicated by ``projected (best-fit)".

\subsubsection{Deprojected spectral analysis}
In addition to a projected spectral analysis, we also carried out a deprojected spectral analysis 
for regions 1--6. 
{\sl XSPEC v12} allows us to treat group and background emissions separately 
in the fitting, with different models and different ARFs: 
we want to deproject only the group emission 
and the background is more uniformly distributed than the group emission.  
Spectra from regions 1--6 were read into {\sl XSPEC} as 6 data groups.  
First, we represented group emission with model 
{\tt projct}$\times$({\tt phabs}$\times$({\tt apec$_{\rm ESO}$}+$\tt pow_{\rm LMXB}$));
the normalization of the LMXB component is a variable fitting parameter in (the central) region 1 and was fixed at 0 for regions 2--6. 
The {\tt projct} model performs a projection of 3-D shells into 2-D annuli with the same center.   
We adjusted the values of the $\beta$-model image ARFs so that they  
corresponded to complete (360$^\circ$) annuli for the source regions of interest (regions 1-6).
These scaled ARFs were used for the group emission (model 1).  
The background emission was represented by a second model:
{\tt apec}$_{\rm LB}+${\tt phabs}$\times$(${\tt apec}$$_{\rm MW}$+${\tt pow_{\rm CXB}}$),
as described above.
Uniform ARFs were used for the background emission (model 2). 
Models 1 and 2 were fitted simultaneously to the spectra. 
To ensure the stability of our deprojection analysis, we needed to tie the 
temperatures and abundances of regions 3 and 4 and, separately, 
the temperatures and abundances of regions 5 and 6. 
We first fixed the surface brightness of each background component as described earlier.  
The associated fitting results are listed in Table~3, labeled ``deprojected (best-fit)."
We next let the surface brightness of each background component freely vary in the fitting. 
In this way, the effects of background statistical uncertainties upon determining 
source properties were taken into account. 
For these fits we performed MCMC simulations with 40 chains,  each chain having 10000 steps.

With the {\tt margin} command in {\it XSPEC}, we marginalized over the 
uncertainties of the normalizations of all source parameters. 
The mean and uncertainty of each fitting parameter was derived from its probability 
distribution and is listed in Table~3, in the block labeled ``deprojected (marginalized)." 
The value of each source parameter is marginalized over all other free source and background parameters. 
Therefore, we adopt these ``deprojected (marginalized)" results as the deprojected results used for further analysis described below in \S3; 
the alternative ``deprojected (best-fit)" results are used only for testing the effects of
systematic uncertainties in each background component, as will be described in \S4.3.

The deprojected electron density of each spherical annulus
was derived from the normalization
of the ${\tt apec}_{\rm ESO}$ component in each deprojected shell obtained.
The normalization of this {\tt apec} model in {\sl XSPEC} is proportional to the emission measure and defined as 
$$ {\rm norm} = \frac{10^{-14}}{4\pi[D_A(1+z)]^2}\int{n_en_H}{dV},$$ 
where $D_A$ is the angular distance to the group, and $V$ is the volume of the spherical annulus. 
We assume that the hot gas density has a single value in the given volume for each spherical annulus.
As indicated by  tests performed by Su \& Irwin (2013) for NGC~720,
when a single volume is divided into 10 spatial bins and analyzed separately, 
the sum of the gas masses of the 10 bins is within 10\% of the gas mass obtained by analyzing a single
integrated bin of the same volume. 
We did not correct for the point spread function (PSF) overlap between adjacent regions caused by the $1.5^{\prime}-2^\prime$ spatial resolution 
of {\sl Suzaku} in our analysis. 
Humphrey et al.\ (2012) found that this effect was insignificant in their results for RXJ~1159+5531
and the annular widths in our work are $\sim50\%$ larger than those of Humphrey et al.\ (2012), 
further reducing such effects.

\subsection {\sl Stellar mass estimates from infrared imaging}

We will be interested in determining the radial profile of the iron mass-to-light ratio in this group.
We used $K$-band imagery from the {\sl WISE} ({\sl Wide-Field Infrared Survey Explorer})
public archive to estimate the stellar light distribution in ESO 3060170.
We analyzed three $2^\circ \times 2^\circ$ images, with one centered on ESO 3060170,
one centered $2^\circ$  east, and one centered $2^\circ$ west of the group center.
All three images were reduced using {\sl SExtractor} to detect point sources.  
Point sources were excluded and replaced with a 
locally interpolated surface brightness using the {\sl CIAO} tool {\tt dmfilth}.   
We determined the surface brightness of the local infrared background using the 
average of the two offset pointings to the east and west, which were consistent
with each other within 2\%. 
We subtracted this average background surface brightness from the infrared
surface brightness in each of the areas subtended by regions 1-6.
We converted source counts in each region to the corresponding magnitude 
and corrected for Galactic extinction.
We adopted a Solar $K$-band absolute magnitude of $M_{K\odot}=3.28$ 
(Binney \& Merrifield 1998) and derived the stellar mass by adopting 
a stellar mass-to-light ratio of $M_*/L_{K}$ of 1.17 $M_{\odot}/L_{K\odot}$ 
(Longhetti et al.\ 2009).

\section{\bf Results}
\smallskip
\medskip

\subsection{\bf Temperature and density}

Projected and deprojected temperature profiles are shown in Figure~\ref{fig:temperature}. 
The projected temperature profile is roughly
isothermal ($\sim 2.5$ keV) within $0.8 R_{200}$ (800 kpc); this is consistent with the temperature profile
between 20 kpc and 400 kpc found by Sun et al.\ (2004) from {\sl Chandra} and
{\sl XMM-Newton} data. 
The temperature drops quickly beyond 800 kpc  to 
$1.13^{+0.36}_{-0.19}$ keV at $R_{\rm 200}$ (1.15 Mpc).  
This temperature drop (by a factor of $\sim 2.2$) is smaller than found on the same
normalized scale in other clusters observed with {\sl Suzaku} (Akamatsu at al.\ 2011).

Temperatures of regions 3 and 4 were linked, as were the temperatures of regions 5 and 6.
The projected and deprojected temperatures of the outer two bins are consistent, so
we combined the deprojected temperatures of the inner 4 bins with the projected temperatures of the outer
two bins to construct a global temperature profile for further analysis. 
This global temperature profile is plotted in Figure~\ref{fig:temperature}, where it is shown to be consistent 
with the following heuristic formula 
calibrated by Pratt et al.\ (2007) for clusters over a radial range of $0.125 < R/R_{200} < 0.5$:
$$ T/T_X=1.19-0.74~R/R_{200}, $$ 
where $T_X$ is the peak temperature.

Higher resolution {\sl Chandra} observations analyzed by Sun et al.\ (2004)
revealed a small cool core of $1.32^{+0.08}_{-0.10}$ keV within 10 kpc (0.24$^\prime$). 
They also found that the gaseous cooling time $t_c$ is less than the Hubble time $t_H$
out to 70 kpc ($1.7^\prime$).
The fact that the cool core is much smaller than the region over which $t_c<t_H$
suggests that the central regions have been heated (Sun et al.\ 2004).
{\sl Suzaku} cannot spatially resolve such a small cool core due to its relatively 
 large PSF. 
Nonetheless, we looked for the signature of this cool core in a 
{\sl Suzaku} spectrum extracted from within a central circular region of $3^\prime$ diameter.
We fitted this central spectrum with two independently varying thermal diffuse emission components, 
along with the LMXB component described \S2.2.
The two temperatures we obtained were $1.03^{+0.63}_{-0.33}$ keV 
and
$2.76^{+0.16}_{-0.16}$ keV, with the lower temperature component being consistent with
the $1.3$ keV found by Sun et al.\ (2004).

The deprojected electron density profile, shown in Figure~\ref{fig:density}, is derived from a combined fit for all annular regions. 
The profile approximately follows a power law model $n_e \propto r^{-\delta}$, with a best fit of $\delta = 2.29 \pm 1.4$. 
We also fit the profile to a $\beta$-model, $n_e=n_{e,0}[1+(r/r_c)^2]^{-3\beta/2}$,
obtaining a best fit of $\beta=0.78\pm0.29$.
\subsection{\bf Entropy and pressure}

The entropy of gas in groups and clusters is a powerful diagnostic for investigating 
gaseous heating processes in such systems. 
Figure~\ref{fig:entropy} shows the entropy 
($K = n_e^{-2/3}kT$) profile of ESO 3060170 derived from these
{\sl Suzaku} observations out to $R_{200}$ (1.15 Mpc). 
Within $R_{500}$ (750 kpc),
the best-fit power law index of this entropy profile is 
$0.81\pm0.23$, which was obtained with the inner four data points. 
This is consistent with the average
entropy profile of 31 clusters observed with {\sl XMM-Newton} (Pratt et al.\ 2010), 
for which a best-fit power law index of 0.98 was found out to about $R_{500}$;
the entropy profile of ESO 3060170 is also 
consistent with those of $\sim20$ galaxy groups observed within $R_{500}$ by Sun et al.\ (2009),
who found an average best-fit power law index of 0.7. 
If purely gravitational processes energize the gas, self-similar 
structure formation simulations predict that the entropy should rise with radius as 
$S \propto r^{1.1}$ (Voit et al.\ 2005). 
In ESO 3060170, the observed entropy profile flattens and declines beyond $\sim R_{\rm 200}$ (1.15 Mpc),
regardless of which normalization was adopted for the expected entropy profile. 
We tested whether this behavior at large radii is an artifact of underestimating the background:
as discussed in detail in \S4.3, we varied each background component and found this entropy rollover to persist at large radii.
A similar entropy downturn has been found in several massive clusters observed at large radii by {\sl Suzaku} 
(Walker et al.\ 2012c).

We would like to compare the observed entropy profile to expectations from
numerical simulations and to those of other groups and clusters. 
We estimated the normalization of the expected entropy profile in two different ways, 
as shown in Figure~\ref{fig:entropy}.
The first estimate, $K_{\rm sim}$, is based on the simulation of gravitational structure formation
derived by Voit et al.\ (2005):

$$K_{\rm sim}(R)=1.32~K_{200}(R/R_{200})^{1.1},$$
where the normalization $K_{200}$ is given by 
$$K_{200}=362~\frac{GM_{200}\mu m_p}{2R_{200}}\left(\frac{1}{\rm keV}\right)\times\left[\frac{H(z)}{H_0}\right]^{-4/3}\left(\frac{\Omega_m}{0.3}\right)^{-4/3} ~{\rm keV ~cm^{-2}}.$$

Our second estimate for the expected entropy profile, $K_{\rm fit}$, is based
on fitting the observed entropy profile of ESO 3060170 between 0.12-0.8 $R_{\rm 200}$;
at these intermediate radii, the gas is likely to be virialized 
and relatively free from a possible central disturbance such as AGN activity. 
The observed entropy profile displays a more extended central excess when compared with $K_{\rm sim}$ rather 
than $K_{\rm fit}$; this was also found by Eckert et al.\ (2013) in an analogous comparison
within a cluster sample. 
In systems with radial variations in the enclosed gas mass fraction $f_{\rm gas}(r)$, 
Pratt et al.\ (2010) suggests that 
the observed entropy profile should be scaled by a correction factor 
$[f_{\rm gas}(r)/0.15]^{2/3}$ when comparing to numerical models. 
We modified this scaling factor to $[f_{\rm gas}(r)/0.15]^{2/3}$, 
based on the new {\sl Planck} result for
the cosmic baryon fraction ($f_b=0.15$; {\sl Planck} Collaboration 2013). 
After applying this correction factor, the scaled entropy profile (blue stars) at intermediate radii 
in Figure~\ref{fig:entropy} becomes more consistent with $K_{\rm sim}$ (red solid line) 
than the original entropy profile (black stars).

The gas pressure ($P=n_e kT$) profile of this fossil group is shown in Figure~\ref{fig:pressure} and
compared to a semi-analytic universal pressure profile derived by Arnaud et al.\ (2010)
from comparison of their numerical simulations 
to {\sl XMM-Newton} observations of clusters.
This pressure profile is characterized as
$$
P(r)=P_{500}\left[\frac{M_{500}}{1.3\times10^{14} h_{70}^{-1}M_{\odot}}\right]^{a_p+a^{\prime}_p(x)}\times\frac{P_0}{(c_{500}x)^{\gamma}[1+(c_{500}x)^{\alpha}]^{(\beta-\gamma)/\alpha}},
$$
where 
$$x=\frac{r}{R_{500}}; ~ ~ ~ ~ ~ a^{\prime}_p(x)=0.10-(a_p+0.10)\frac{(x/0.5)^3}{[1+(x/0.5)^3]};$$ 
and $P_{500}$ and $M_{500}$ are respectively the pressure and total mass at $R_{500}$. 
Arnaud et al.\ (2010) adopted parameter values of 
$$[P_0, ~c_{500}, ~\gamma, ~\alpha, ~\beta] ~=~ [8.403h_{70}^{-3/2}, ~1.177, ~0.3081, ~1.0510, ~5.4905].$$
The observed pressure is consistent with the universal profile  within 0.8$R_{200}$,
while it exceeds the universal profile at large radii.

\subsection{\bf Metal abundances}

The deprojected iron abundance profile for this 
fossil group is listed in Table~3 and shown in Figure~\ref{fig:abundance}. 
To ensure the stability of our deprojection analysis, we needed to tie the 
temperatures and abundances of regions 3 and 4 and, 
separately, the temperatures and abundances of regions 5 and 6.
The best-fit iron abundance at $R_{200}$ (1.05 Mpc; 27.5$^\prime$) is $0.25\pm0.20$ $Z_\odot$.
Note that possibly clumpy gas would imply multi-phase thermal structure, 
which can lead to underestimates of the iron abundance (Buote 2002). 
We tested for the presence of multiphase gas in the outer parts
of ESO 3060170 by fitting a spectral model with two temperature components;
the fit was not improved compared to the single temperature model.
However, further {\it XSPEC} analysis of synthetic two-temperature models
showed that the spectrum in the
outer parts is sufficiently background-dominated that it would be very difficult
to separate two plausibly distinct temperature components.
As indicated above, the best-fit iron abundance at 
$R_{200}$ is 0.25 Z$_\odot$ (corresponding to 0.17 Z$_\odot$ for
Anders \& Grevesse (1989) solar abundances).
This is in line with
previous results for massive clusters such as 
Virgo ($Z_{\rm Fe} \gtrsim 0.1~Z_{\odot}$, Urban et al.\ 2011),
Perseus ($Z_{\rm Fe}\approx0.3$ Z$_{\odot}$, Simionescu et al.\ 2011),
A1689 ($Z_{\rm Fe}=0.35^{+0.65}_{-0.31}$ Z$_{\odot}$, Kawaharada et al.\ 2010), 
and A399/401 ($Z_{\rm Fe}\approx0.2$ Z$_\odot$, Fujita et al.\ 2008).
Such possibly significant iron abundance at such large radii could be indicative of
early enrichment, ram-pressure stripping of infalling galaxies,
or redistributed enriched gas from more central regions of the cluster.  

The iron-mass-to-light ratio (IMLR) tends to increase with radius in groups and clusters 
(Sato et al.\ 2012), implying that the distribution of metals is usually more
extended than the galaxy light. 
To determine the IMLR profile in ESO 3060170, we derived the iron mass distribution from
the deprojected profiles of iron abundance and gas mass (see \S3.4); 
the stellar light distribution was derived from 
the $K$-band imagery described above.
In Figure~\ref{fig:IMLR} we show the enclosed IMLR profile and compare it to several clusters. 
There is a large scatter among these IMLR profiles; these systems have a range of masses 
and temperatures and probably have different formation epochs and enrichment histories.
Still, we note that  
the enclosed IMLR of ESO 3060170 and the NGC~5044 group are similar in their inner regions; 
both groups have significantly smaller IMLR values than the other plotted systems 
beyond $R/R_{200}\approx0.1$, likely for different reasons.  
NGC~5044 is a typical galaxy group with a temperature of $\sim1$ keV. 
Groups with shallower potentials may have their enriched gas redistributed out to large radii since they are more vulnerable to AGN feedback and galactic winds. 
In addition, galaxy groups tend to have larger stellar mass fractions than galaxy clusters, which may further reduce their IMLRs.  
As for ESO 3060170, fossil groups tend to have more concentrated stellar luminosity profiles than other groups,
as a natural consequence of the defining characteristic of fossil groups. 
Thus, ESO 3060170 has a relatively smaller IMLR towards the group center.
Its enclosed IMLR  becomes comparable to cluster values ($> 1.0\times10^{-3}$ $M_{\odot}/L_{K\odot}$) by $R_{\rm 500}$,
where IMLR $\sim3\times10^{-3}$ $M_{\odot}/L_{K\odot}$;
this is much larger than 
is typical for groups, which have IMLR values ranging from $\sim10^{-5}-10^{-3}~M_{\odot}/L_{K\odot}$ (Renzini 1997),
although most other such groups (such as NGC~5044) have yet to be observed to their virial radii.

\subsection{\bf Mass}
The temperature and density profiles were used to calculate the total mass distribution 
from the equation of hydrostatic equilibrium:
$$ M(r)=-\frac{kT(r)r}{\mu m_p G}\left(\frac{{\rm d~ln~\rho_{g}(r)}}{{\rm d~ln}~r}+
\frac{{\rm d~ln}~T(r)}{{\rm d~ln}~r}\right). $$
The observed global temperature profile (described in \S3.1) was fitted with a three-dimensional 
(deprojected) temperature profile characterized by Vikhilinin et al.\ (2006): 
$$T(r)=\frac{T_0(r/r_{\rm cool})^{a_{\rm cool}}+T_{\rm min}}{(r/r_{\rm cool})^{a_{\rm cool}}+1}\times\frac{(r/r_t)^{-a}}{[1+(r/r_t)^b]^{c/b}}.$$
Uncertainties in mass estimates were assessed using Monte Carlo realizations
of the temperature and density profiles. 
Total mass and gas mass profiles are shown in Figure~\ref{fig:mass}; the enclosed gas 
mass profile and the enclosed baryon mass fraction profile are shown in Figure~\ref{fig:fg}. 
The total mass within $R_{\rm 200}$ (1.15 Mpc) is estimated to be $1.7\times10^{14}M_{\odot}$;
this leads to a gas mass fraction of $f_{\rm gas}=0.11^{+0.02}_{-0.02}$. 
Including the stellar
mass contribution as described in \S2.3, we obtained  
a baryon fraction of $f_{b}\approx0.13^{+0.02}_{-0.02}$, consistent with the 
cosmological baryon fraction $f_b=0.15$ ({\sl Planck} Collaboration 2013) and comparable to those of 
some massive clusters (cf.\ George et al.\ 2009; Kawaharada et al.\ 2010).
In Figure~\ref{fig:mass} we also plot the total mass and gas mass profiles derived by 
Sun et al.\ (2004) from {\sl Chandra} and {\sl XMM-Newton} 
observations of ESO 3060170 within 450 kpc; 
our results are consistent within the regions of overlap. 
Our total mass is also in line with the extrapolation by Sun et al.\ (2004) to $R_{\rm vir}$. 

We obtained the dark matter distribution in this group by subtracting the baryonic mass 
distribution from the total mass distribution. 
In Figure~\ref{fig:nfw} we compare the resulting  dark matter mass profile to the  
quasi-universal  dark matter mass profile of Navarro et al.\ (1997):
$$M(r)=4\pi  \delta_{c}\rho_c(z){r_s}^3m(r/r_s),$$
where
$$m(x)={\rm ln}(1+x)-\frac{x}{1+x}$$
and the associated dark matter density profile is
$$ \rho(r)=\frac{\rho_{c}(z)\delta_{c}}{(r/r_{s})(1+r/r_{s})^{2}}; $$
here 
$r_{s}$ is a scaling radius, 
$ \rho_{c}(z)=3H(z)^{2}/8\pi G $,  and
$$ \delta_{c}=\frac{200}{3}\frac{c^3}{{\rm ln}(1+c)-c/(1+c)}, $$
from which we can determine the dark matter concentration $c$. 
We derived values of $c=6.9\pm 4.5$ and $r_{s}=163\pm121$ kpc.  
Given the large uncertainty, 
such a dark matter concentration is consistent with those of non-fossil systems with similar masses 
and temperatures (Ettori et al.\ 2010).
The derived virial radius  $R_{\rm vir}$=1.12 Mpc ($c\equiv$ $R_{\rm vir}/r_s$) is nearly equal to 
$R_{200}$ (1.15 Mpc).

\section{\bf Discussion}

\subsection{Comparison of density and temperature profiles to simulations}

As shown in \S3.2, our {\sl Suzaku} observations of ESO 3060170 indicate that its entropy
($K=n_{e}^{-2/3}kT$) profile flattens at large radii relative to the 
$K\propto r^{1.1}$ behavior expected from simulations. 
This entropy flattening has also been found in some massive clusters 
(cf.\ Walker et al.\ 2012c).
Currently, there are several proposed explanations for such shallow entropy profiles at large radii. One suggestion is that
the gas is clumpy at large radii (Simionescu et al.\ 2011), 
causing the emissivity-weighted density to be higher than the actual average density. 
An alternative explanation is that 
the ion temperature may be higher than the electron temperature in the outer
parts of a cluster after relatively recent heating by an accretion shock  (Hoshino et al.\ 2010);
thus, the measured electron temperature is lower than the eventual equilibrium 
temperature in these regions. 
Yet another suggestion is due to Cavaliere et al.\ (2011), who propose a cluster evolutionary model incorporating the effects of the weakening of accretion shocks
over time; Walker et al.\ (2012c) finds this latter model is consistent with the observed entropy 
flattening.
It is also possible that more than one of these processes may be occurring. 

In Figure~\ref{fig:entropy} we compared the observed entropy profile (black dots) to the
theoretical predictions $K_{\rm sim}$ (solid red line) from Voit et al.\ (2005) and
in Figure~\ref{fig:pressure} we compared the observed pressure profile (black dots) to the 
theoretical expectations of Arnaud et al.\ (2010).
We derive similar comparisons for the gas density and temperature profiles by
using the following equations for observed and theoretical quantities: 
$$ K_{\rm obs}\propto n_{\rm obs}^{-2/3}T_{\rm obs}; ~ ~ ~ ~
  P_{\rm obs}\propto n_{\rm obs}T_{\rm obs}; ~ ~ ~ ~
 K_{\rm th}\propto n_{\rm th}^{-2/3}T_{\rm th}; ~ ~ ~ ~
P_{\rm th}\propto n_{\rm th}T_{\rm th}.
$$

The following relations are equivalent:
\begin{equation}         
\frac{n_{\rm obs}}{n_{\rm th}}=\left(\frac{P_{\rm obs}K_{\rm th}}{K_{\rm obs}P_{\rm th}}\right)^{3/5}; ~ ~ ~ ~ ~
\frac{T_{\rm obs}}{T_{\rm th}}=\left(\frac{P_{\rm obs}}{P_{\rm th}}\right)^{2/5}\left(\frac{K_{\rm obs}}{K_{\rm th}}\right)^{3/5}.           
\end{equation}         
We used these relations to obtain the ratios of the observed to theoretical 
electron density and temperature profiles shown in Figure~\ref{fig:nt};
adopted normalizations were described in $\S3.2$.

When $K_{\rm sim}$ is adopted for the expected entropy profile,
the observed temperature profile is hotter than expectations throughout 
(see Figure~\ref{fig:nt}a). 
The observed density profile is lower than expectations
in the central regions but is in agreement with expectations at intermediate radii; 
at large radii, the observed density becomes twice as large as expected 
(see Figure~\ref{fig:nt}a).
When $K_{\rm fit}$ is adopted for the expected entropy profile,
the disparity between observations and theoretical expectations is smaller for the 
temperature profile than for the density profile (see Figure~\ref{fig:nt}b).
The temperature profile departs 
by only
$\sim20\%$ in the outer parts, while the density profile departs by more than
a factor of $\gtrsim 2$.
Both results imply that it is the density behavior which is driving the flattening
of the entropy profile relative to theoretical expectations.

The electron density profile being shallower than theoretical expectations
may be due to clumpy gas,
since the emissivity scales as $\langle n^2 \rangle$.
Simulations by Nagai et al.\ (2011) show that gas clumpiness could cause the 
average density at the virial radius to be overestimated by up to a factor of 10, 
with 30\% enhancements more likely.
{\sl Suzaku} observations of the Perseus cluster indicate a gas mass 
fraction of $f_g\approx0.23$,
substantially greater than the cosmic baryon fraction of $f_b=0.15$ 
({\sl Planck} Collaboration 2013), 
assuming the gas is smoothly distributed (Simionescu et al.\ 2011).  
If instead the gas in the outer parts of Perseus is clumpy (which enhances the X-ray emissivity 
for a given average gas density), the actual gas mass fraction may
be less.  
If the true gas mass fraction in Perseus is no more than the cosmic baryon 
fraction, the gas density must be overestimated by a 
factor of 3 at its virial radius  (Simionescu et al.\ 2011).

If the gas in the outer parts of ESO 3060170 is clumpy, 
the temperature may also be underestimated: 
in pressure equilibrium, higher density regions would have cooler temperatures
and higher emissivities per unit volume 
than lower density regions and may dominate the 
emissivity-weighted temperature measures.
As mentioned above, a two-temperature model for the outer regions
did not provide improved fits over single-temperature models;
however, the outer parts are sufficiently background-dominated that 
a multi-temperature model would be difficult to constrain.
The enclosed gas mass fraction in this group increases smoothly with radius and 
never exceeds the cosmic mean.
Thus, the enhancement of the gas density at large radii (relative to theoretical
expectations) may not be simply a result of clumping. 
Instead (or additionally), the gas may have been redistributed outward
due to central heating;
this is also suggested by the gas density being lower 
than theoretical expectations within $0.5R_{\rm 200}$. 
Using a sample of massive clusters observed with {\sl XMM-Newton}, Pratt et al.\ (2010) 
showed that clusters with entropy enhancements within $R_{500}$ 
(relative to theoretical expectations) had lower gas mass fractions. 
In simulating clusters with masses $4-11\times10^{14}M_\odot$,  Mathews \& Guo (2011) 
showed that feedback energy from central black holes far exceeds 
radiative energy losses and causes gas to expand outward beyond the virial radii. 
Fossil group ESO 3060170 has a 
shallower gravitational potential than massive clusters,
so it should be even more vulnerable to such feedback. 
Figure~\ref{fig:entropy} shows that its entropy profile is in good agreement 
with $K_{\rm sim}$ (at intermediate radii) 
after it has been scaled downward by $[f_{\rm gas}(r)/0.15]^{2/3}$ (see \S3.2);
this is consistent with the effects of gas redistribution in shaping the entropy profile.     
Furthermore, the IMLR increases with radius in groups and clusters 
(Sato et al.\ 2012), so the metal distribution is more
extended than the galaxy light, which may be a result of gas expansion. 

We further address this issue quantitatively by comparing 
the absolute amount of gas ``missing" 
in the central regions to the ``excess" 
of gas in the outer regions, both relative to theoretical expectations 
derived from the reference entropy profiles described above.
First, we obtained the theoretically expected density in each region by 
dividing the observed density by the observed-theoretical density ratio derived in eq.(1). 
Thus, we obtained the theoretically expected gas mass of each region. 
We next compare the ``excess" gas mass in the outermost bin to its total ``deficit" in the inner two bins.
If we first use the best-fit entropy profile ($K_{\rm fit}$) described in \S3.2 
as the baseline expectation,
the gas density deficiency in the central regions (red diamonds in Figure~\ref{fig:nt}b) can only provide  
$\sim 46\%$ of the mass associated with the gas density excess at large radii. 
If we assume the other $\sim 54\%$ of the apparent gas mass ``excess" at large radii 
is an artifact of clumpy gas,
the true gas mass of the outermost bin is 80\% of its apparently observed values. 
We would then obtain a total gas mass fraction of $f_{\rm gas}\approx0.105$ 
after correcting for this clumpiness. 
This value is much smaller than the cosmological baryon fraction $f_b\approx 0.15$
(as well as the baryon fraction of $f_b\sim 0.17\pm0.02$ found in
fossil group RXJ 1159+5531 by Humphrey et al.\ 2012).
If we instead use the entropy profile $K_{\rm sim}$ derived from simulations (Voit et al.\ 2005) 
as the baseline expectation,
the gas mass ``missing" in the central regions 
(red solid squares in Figure~\ref{fig:nt}a) can account for $160\%$ of the ``excess" gas mass at larger radii. 
In other words, gas redistribution alone would be sufficient to explain our observational results 
and there would be no need to invoke clumpy gas in the outer parts of this fossil group. 
Moreover, the ``extra" 60\% of gas mass in the central regions would suggest that 10\% of the total gas has been driven out beyond the virial radius
of this group, perhaps by central AGN activity or galactic winds;
this would also be consistent with the enclosed baryon fraction of this 
fossil group/poor cluster being somewhat smaller than the cosmological value.
Another possibility is that the 60\% deficiency of gas mass in the central regions 
may have been consumed by star formation.   
Without any more knowledge such as which reference entropy profile is more appropriate 
to be adopted, we conservatively conclude that
the flattening of the entropy profile at large radii in this fossil group may be due to a 
combination of multiple factors including gas clumpiness and/or gas redistribution.
We note that there are large uncertainties in the gas density excesses and deficits derived
from respectively adopting the $K_{\rm sim}$ and $K_{\rm fit}$ reference entropy profiles;
consequently, we cannot make a statistically significant distinction
between the clumpiness and redistribution scenarios.

\subsection{The nature of fossil groups}

The entropy profile of fossil group ESO 3060170 is observed to rise outward then roll over
at $0.9R_{200}$; this is consistent with the entropy behavior in several massive clusters
observed with {\sl Suzaku}, but contrary to that of 
fossil group RXJ 1159+5531
 (Humphrey et al.\ 2012).
Most {\sl Suzaku} observations of clusters have  extended to only a fraction of their virial radii,
therefore covering only a small fraction of the circular projected area within $R_{\rm vir}$.
It has been suggested that
azimuthal variations in cluster properties may 
be responsible for the observed scatter in {\sl Suzaku} observations of
cluster entropy profiles at large radii 
(Eckert et al.\ 2012; Walker et al.\ 2012c). 
Azimuthally averaged studies of massive clusters with the {\sl ROSAT} and {\sl Planck} satellites
find entropy profiles that rise steadily with radius (Eckert et al.\ 2013), in contrast to the entropy rollovers 
found in {\sl Suzaku} cluster observations with more limited azimuthal coverage.
The Humphrey et al.\ (2012) study of fossil group RXJ 1159+5531 was also azimuthally more complete
than our study of ESO 3060170, since RXJ 1159+5531 is more distant and has a smaller
temperature (thus, a smaller virial radius).
We would need more complete areal coverage of ESO 3060170 to test whether 
its entropy profile varies azimuthally.
Alternatively, these disparate entropy profiles may indicate that fossil groups are 
rather heterogeneous, with various evolutionary histories.

Simulations  predict that fossil systems tend to reside in less dense environments 
than non-fossil systems with similar masses and formation epochs 
(D\'iaz-Gim\'enez et al.\ 2011; D'Onghia 2005; Dariush 2007).
We compared the environment of fossil group ESO~3060170 
($kT = 2.8$ keV; $D_L$ = 154 Mpc)
to that of the normal poor cluster Hydra~A, which has a similar gas temperature 
($kT = 3.0$ keV) and a 50\% larger distance ($D_L$ = 238 Mpc).
Both systems are in the southern sky and are within the coverage of the 
6dF galaxy redshift survey (Jones et al.\ 2009), from which we selected
galaxies within a (25 Mpc)$^3$ cube centered on each system.
We only included galaxies with 
reliably determined redshifts and selected galaxies to the same limiting
luminosity in each system.  The 6dF sample is nearly complete to an $R$ magnitude
of 15.6, so we selected galaxies to this magnitude limit around the more distant 
Hydra~A system; around ESO 3060170 we selected galaxies to the equivalent 
limiting luminosity, corresponding to a magnitude limit 0.94 mag brighter.
There are 82 such galaxies in the (25 Mpc)$^3$ volume around Hydra~A,
with a total $R$-band luminosity of 2.76$\times10^{12}$ $L_{\odot}$. 
There are only 20 such galaxies in the (25 Mpc)$^3$ volume 
around ESO~3060170, with a total $R$-band luminosity of 6.3$\times10^{11}$ $L_{\odot}$
(less than a quarter of that around Hydra~A).
In Figure~\ref{fig:map} we show galaxy maps on a scale of 25 Mpc centered 
on each system. 
Hydra~A is connected to rich cluster Abell~754 (to the northwest) by 
a structure described as a filament by Sato et al.\ (2012), 
while ESO~3060170 resides in a more isolated environment. 
A $B$-band comparison yielded similar results.

Most fossil groups have larger mass-to-light ratios than non-fossil groups at a 
given total mass, implying that fossil groups are optically deficient 
(Proctor et al.\ 2011).
Jones et al.\  (2003) noted that fossil groups with temperatures comparable to Virgo 
($\sim2$ keV) lack galaxies more luminous than $L_*$ (the characteristic luminosity of the 
Schechter luminosity function; Schechter et al.\ 1976), whereas the Virgo cluster contains six $L_*$ galaxies (Jones et al.\  2003). 
ESO~3060170 is also optically deficient ($M/L_{R}\sim 230~ M_{\odot}/L_{R\odot}$),
as is typical of fossil groups (Figure~4 in Proctor et al.\ 2011).  
This may be a consequence of 
relatively inefficient star/galaxy formation.

\subsection{Systematic uncertainties}
We presented  {\sl Suzaku} observations of the hot gas in fossil group ESO 3060170 out to its virial radius. 
While there are numerous X-ray studies of massive clusters to their virial radii (cited in \S1),
there are relatively few studies of groups to such radii.
This is largely due to the relatively low X-ray surface brightness of groups, 
which also makes them more vulnerable to systematic uncertainties.
In order to explore how systematic uncertainties in the background can affect 
our results and conclusions, we varied the 
adopted surface brightness of each component
of the X-ray background (CXB, MW, LB) by $\pm$20\% and we varied the count rate of the 
particle background (NXB) by $\pm$5\% in successive deprojected spectral fits 
adopting fixed  background components.
The effects of these background variations on the entropy profile are shown in Figure~\ref{fig:entropy_sys}. 
The derived entropy profile at large radii, its rollover in particular, 
is robust against these various systematic uncertainties.  
The impact of these background variations on our results in the outermost region
are shown in Table~4: 
for comparison, we first list the marginalized values for gas properties
(as well as enclosed mass, gas mass fraction, and dark matter concentration)
derived from fits with freely varying background components (in the row labeled ``Marginalized");
we next list the best-fit values derived from fits adopting fixed background components
(in the row labeled ``best-fit");
these latter fitting results with fixed background components are the 
baseline for the systematic
background variations explored in the remainder of Table~4.
For each background variation, the rest of Table~4 lists
how these derived quantities differ from those of the initial best-fit model.
It is worth noting that we assume that each X-ray background component 
has the same surface brightness 
in all regions, which is a conventional assumption but may not be the case. 
The surface brightness fluctuations of unresolved cosmic point sources in the 0.5-2.0 keV band
are expected to be 
$\sigma_B=3.9\times10^{-12} \Omega^{-1/2}_{0.01}$erg s$^{-1}$ cm$^{-2}$ deg$^{-2}$
(where $\Omega_{0.01}$ is the solid angle of the target region in units of 0.01 deg$^{2}$; Bautz et al.\ 2009); 
this implies that the surface brightness fluctuations of unresolved point sources would be
$\sim 5\times10^{-14}$ erg s$^{-1}$cm$^{-2}$ in the outermost region. 
To constrain such CXB variations is beyond the ability of current {\sl Suzaku} observations. 
In addition, there are many other systematic uncertainties such as solar wind charge exchange, 
stray light, PSF, stellar mass, distance, etc.\ that we have not explored in our study. 
Humphrey et al.\ (2012) shows that these latter effects are relatively secondary for their 
{\sl Suzaku} observation of RXJ~1159+5531. 
ESO 3060170 is closer and more massive than RXJ~1159+5531, 
so our results should be even less sensitive to these secondary factors.

Inspection of Table 4 and Figure~\ref{fig:entropy_sys} shows that variations in
the CXB component produce the 
largest changes in our outermost entropy value (by $14-24$\%).
Consequently, we explored the systematic uncertainties associated with
the CXB modeling itself.
As described in \S2.2.1, we adopted a single power law model 
for the CXB ($\Gamma=1.41$; De Luca \& Molendi 2004) in our spectral analysis.
Alternatively, we considered another commonly used model for the CXB (Yoshino et al.\ 2009) 
consisting of two broken power laws (Smith et al.\ 2007). 
These two components 
represent the integrated spectra of bright and faint AGNs separately.
The bright AGNs were modeled with a photon index of $\Gamma=1.4$ above a break energy of 
1.2 keV, steepening to $\Gamma = 1.96$ for energies
below 1.2 keV;
its normalization was allowed to vary 
(but was constrained to be less than 4.90 photons s$^{-1}$ cm$^{-2}$ sr$^{-1}$  at 1 keV;
Smith et al.\ 2007).
Faint AGNs were also modeled with $\Gamma=1.4$ above a break energy of 
1.2 keV, but steepening to $\Gamma = 1.54$ for energies
below 1.2 keV.
This flatter power law  
has been well determined by many observations. 
The integrated spectrum of faint AGNs has smaller statistical fluctuations, so
its normalization was fixed (at 5.70 photons s$^{-1}$ cm$^{-2}$ sr$^{-1}$  at 1 keV;
Smith et al.\ 2007; Yoshino et al.\ 2009).  
We adopted this dual broken power law model ({\tt bknpow}$_{\rm bright}$+{\tt bknpow}$_{\rm faint}$) 
instead of the prior single power law  model ({\tt pow}$_{\rm CXB}$, $\Gamma=1.41$) for the CXB
and repeated the background determination and deprojected spectral analysis described in \S2.2. 
The best-fit results for this background determination are listed in Table~2 
(labeled ``Fitting CXB-2"). 
When we marginalized over the uncertainties in the bright AGN normalization,
we found two peaks in its probability distribution (see Figure~\ref{fig:margin}).
The normalizations associated with both peaks are within the scatter of those measured 
by Yoshino et al.\ (2009) in their analysis of 14 {\sl Suzaku} background fields,
so we adopted each, in turn, and repeated our spectral analysis.
We list in Table~2 the normalization (and associated surface brightness) determined by this 
marginalization for each background component,
including each of the two marginalized values found for the bright AGN component
(labeled ``CXB-2 low" and ``CXB-2 high").
We list in Table~4 the effects on our results after adopting each of the two bright AGN normalizations, in turn
(lines labeled ``CXB-2 low" and ``CXB-2 high");
we plot their associated entropy profiles in Figure~\ref{fig:entropy_sys} (with the same labels).
These results are largely in line with our prior results, 
although Table~4 shows that adopting the higher bright AGN normalization (CXB-2 high) 
leads to a somewhat smaller gas density at large radii. 
Since these data are background-dominated near the virial radius,
we are not able to vary the normalizations of 
both broken power law components in order to fully constrain this CXB model.

\section{\bf Summary}
We observed the X-ray brightest fossil group ESO 3060170 with {\sl Suzaku}
out to $R_{200}$ (1.15 Mpc or $27.5^{\prime}$).  
The gas and baryon mass fractions in this group are larger than typical 
for groups observed within $R_{\rm 500}$, but comparable to cluster values.
This observation suggests that 
the discrepancy between groups and clusters in their gas and baryon mass fractions
will be reduced if groups are observed to sufficiently large radii.  
The integrated iron-mass-to-light ratio of this group is also larger than is 
typical for groups and comparable to cluster values. 
The entropy profile within $R_{500}$ (750 kpc) rises outward as $r^{1.1}$, in 
agreement with theoretical simulations incorporating purely gravitational processes.
However, at larger radii, in the vicinity of $0.9R_{\rm 200}$, the entropy profile 
flattens and starts to decline outward. 
This flattening is primarily due to the gas density profile being 
shallower than theoretical expectations, possibly due to a 
combination of gas clumpiness and outward redistribution.
This group has a high mass-to-light ratio, thus is optically deficient, 
as is typical for fossil groups. 
This may be a consequence of relatively inefficient star/galaxy formation,
as perhaps indicated by a galaxy luminosity surface density map of the 
neighborhood within 25 Mpc of  ESO 3060170.
Our results are largely robust against plausible variations in
the adopted background components.
Further observations of groups at and beyond their virial radii 
are needed to more clearly establish the relationship between clusters and groups 
in the local Universe.

\section{Acknowledgments}
We would like to thank Liyi Gu, Jimmy Irwin, Evan Million, and Ka-Wah Wong for useful discussions and suggestions. We thank Kyoko Matsushita for providing data for Figure~\ref{fig:IMLR}. Y.S.\ wishes to thank Steve Allen for hospitality during her visit to Stanford University. 
We gratefully acknowledge partial support from NASA {\sl Suzaku} grant NNX10AR33G. 
Y.S.\ has been partially supported by a University of Alabama Graduate Council Research Fellowship.

\clearpage

\begin{table}
\caption{Observation log}
\begin{tabular}{lccccc}
\hline
Name& Obs ID & Effective Exposure & R.A. & Dec. & Dectectors\\

\hline
Center&805075010 & 27 ksec & 05h 40m 07.20s & -40d 57m 36.0s & XIS0,1,3 \\
$17^{\prime}$ Offset & 805075060 & 70 ksec & 05h 40m 07.20s  &-41d 15m 00.0s&XIS0,1,3 \\
\hline
\end{tabular}
\end{table}

\begin{table}
\caption{X-ray background determinations}
\begin{tabular}{lllllllc}

\hline
 & \multicolumn{2}{c}{Cosmic} & \multicolumn{2}{c}{Milky Way} & \multicolumn{2}{c}{Local Bubble} & \\
Method&$S_{\rm CXB}^{a}$&Norm$^{b}$&$S_{\rm MW}^{c}$&Norm$^{d}$&$S_{\rm LB}^{c}$&Norm$^{d}$&$C/$d.o.f\\
\hline
Offset  &$14.4^{+1.3}_{-1.0}$&$6.3^{+0.6}_{-0.4}$&$9.7^{+2.8}_{-2.8}$&$5.8^{+1.7}_{-1.7}$ &\phn$7.7^{+1.2}_{-1.2}$&$43.2^{+6.8}_{-6.5}$& 2484/3019\\
Local &$14.8^{+1.1}_{-1.1}$&$6.5^{+0.5}_{-0.5}$&$9.1^{+1.9}_{-1.8}$ &$5.4^{+1.1}_{-1.1}$&$15.3^{+2.4}_{-2.4}$&$83.2^{+12.9}_{-12.7}$&12764/14580\\
Deproj &$14.8^{+1.2}_{-1.1}$&$6.5^{+0.5}_{-0.5}$&$8.6^{+2.3}_{-2.3}$ &$5.2^{+1.4}_{-1.4}$&$15.6^{+2.5}_{-2.4}$&$84.8^{+13.5}_{-12.8}$&16267/18350\\
Deproj $^{\ast}$ &$14.8^{+1.6}_{-1.6}$&$6.5^{+0.7}_{-0.7}$&$9.0^{+2.9}_{-2.9}$ &$5.4^{+1.7}_{-1.7}$&$15.0^{+3.3}_{-3.3}$&$81.5^{+17.9}_{-17.9}$&\\
\hline
Fitting CXB-2 &${15.7^{+16.3}_{-4.2}}^{e}$&${1.3^{+1.3}_{-0.3}}^{f}$&$7.8^{+3.1}_{-5.4}$ &$4.7^{+1.9}_{-3.2}$&$15.3^{+2.1}_{-7.8}$&$83.3^{+11.5}_{-42.5}$&16266/18350\\
CXB-2 low$^{\ast}$ &${21.3^{+9.3}_{-9.3}}^{e}$&${1.6^{+0.7}_{-0.7}}^f$&\multirow{2}{*}{$\Big\}$$9.1^{+2.8}_{-2.8}$} &\multirow{2}{*}{$\Big\}$$5.5^{+1.7}_{-1.7}$}&\multirow{2}{*}{$\Big\}$$14.6^{+3.4}_{-3.4}$}&\multirow{2}{*}{$\Big\}$$79.6^{+18.4}_{-18.4}$}&\\
CXB-2 high$^{\ast}$ &${{31.9}^{+4.0}_{-4.0}}^e$&${2.4^{+0.3}_{-0.3}}^f$&$$&$$&$$&$$&\\
\hline
\tablecomments
{
 \\
a) Surface brightness of CXB ($0.5-2.0$ keV) 
in units of $10^{-9}$ ergs s$^{-1}$ cm$^{-2}$ sr$^{-1}$. \\
b) Normalization of single power law component of CXB (with photon index 1.41), in units of  photons s$^{-1}$ cm$^{-2}$ keV$^{-1}$ sr$^{-1}$ at 1 keV.  \\
c) Surface brightness of Milky Way background ($0.5-2.0$ keV)
in units of $10^{-9}$ ergs s$^{-1}$ cm$^{-2}$ sr$^{-1}$;
$S_{\rm MW}$ uncertainties are proportional to those of their associated normalizations. \\
d) Normalization of Milky Way background is the integrated line-of-sight emission measure, 
$(1/4\pi)\int n_e n_H ds$, in units of $10^{14}$ cm$^{-5}$ sr$^{-1}$.\\
e) Surface brightness of Local Bubble background (0.5-2.0 keV) in units of
$10^{-9}$ ergs s$^{-1}$ cm$^{-2}$ sr$^{-1}$, 
which includes the contribution from both broken power law components.\\
f) Normalization of broken power law component with photon index of 1.96 below 1.2 keV, 
in units of photons s$^{-1}$ cm$^{-2}$ keV$^{-1}$ sr$^{-1}$ at 1 keV.  
The normalization of broken power law component with photon index of 1.54 below 1.2 keV 
was fixed at 5.7 photons s$^{-1}$ cm$^{-2}$ keV$^{-1}$ sr$^{-1}$ at 1 keV.\\
$^\ast$ marginalized values.\\
}
\end{tabular}
\end{table}

\begin{table}
\caption{Summary of fit parameters for group emission in regions 1-6}
\small{
\begin{tabular}{ccccrrr}
\hline
Method & Annuli & Temperature & Fe & $\rm Norm^{a}$ & ${S_{X}}^{b}$ & $C$/d.o.f \\
 & (arcmin) &(keV) &($Z_{\odot}$) & & &\\

\hline
projected & 0--3  &$2.51^{+0.11}_{-0.12}$&$0.69^{+0.08}_{-0.07}$ &$2037.04^{+129.63}_{-120.37}$&$1231.48^{+78.36}_{-72.77}$\phn& 3506/3770\\
(best-fit)&3--8  &$2.72^{+0.14}_{-0.08}$&$0.39^{+0.08}_{-0.07}$ &$425.93^{+18.52}_{-18.52}$\phn&$228.70^{+9.94}_{-9.94}$\phn\phn& 3341/3669\\
&8--14  &$2.37^{+0.22}_{-0.17}$&$0.25^{+0.12}_{-0.09}$ &$124.08^{+9.26}_{-9.26}$\phn\phn&$63.33^{+4.72}_{-4.72}$\phn\phn& 3075/3307\\
&14--20 &$2.42^{+0.85}_{-0.65}$&$0.18^{+0.40}_{-0.17}$ &$35.09^{+5.56}_{-7.41}$\phn\phn&$17.52^{+2.78}_{-3.67}$\phn\phn& 1714/2082\\
&20--25  &$1.43^{+0.55}_{-0.21}$&$0.34^{+1.15}_{-0.29}$ &$9.81^{+2.22}_{-2.04}$\phn\phn&$5.12^{+1.16}_{-1.06}$\phn\phn& 2454/2849\\
&25--30  &$1.13^{+0.36}_{-0.19}$&$-$ &$6.47^{+2.13}_{-1.85}$\phn\phn&$3.92^{+1.29}_{-1.12}$\phn\phn&2179/2676\\
\hline
\hline
\\
deprojected & 0--3  &$2.47^{+0.14}_{-0.16}$&$0.81^{+0.13}_{-0.11}$ &$1438.66^{+129.65}_{-129.65}$&$1263.01^{+113.82}_{-113.82}$& \multirow{6}{*}{16266/18353}\\
(best-fit)&3--8  &$2.82^{+0.15}_{-0.15}$&$0.43^{+0.10}_{-0.09}$ &$435.93^{+19.85}_{-19.16}$\phn&$234.05^{+10.66}_{-10.29}$\phn& \\
&8--14  &\multirow{2}{*}{$\Big\}$$2.48^{+0.21}_{-0.20}$$^{\ast}$\hspace{3 pt}}&\multirow{2}{*}{$\Big\}$$0.25^{+0.13}_{-0.10}$$^{\ast}$\hspace{3 pt}}&$124.32^{+11.12}_{-10.55}$\phn&$64.16^{+5.74}_{-5.44}$\phn\phn& \\
&14--20 &&&$49.26^{+8.12}_{-7.75}$\phn\phn&$17.77^{+2.93}_{-2.80}$\phn\phn& \\
&20--25  &\multirow{2}{*}{$\Big\}$$1.27^{+0.24}_{-0.21}$$^{\ast}$\hspace{3 pt}}&\multirow{2}{*}
{$\Big\}$$0.31^{+0.44}_{-0.20}$$^{\ast}$\hspace{3 pt}}&$7.51^{+5.86}_{-5.69}$\phn\phn&$5.22^{+4.07}_{-3.95}$\phn\phn& \\
&25--30  &&&$17.79^{+4.93}_{-4.79}$\phn\phn&$3.89^{+1.08}_{-1.05}$\phn\phn& \\
\hline
\\
deprojected & 0--3  &$2.46\pm0.20$&$0.81\pm0.14$ &$1421.94\pm150.56$&$1248.41\pm132.18$& \multirow{6}{*}{16266/18350}\\
(marginalized)&3--8  &$2.82\pm0.20$&$0.43\pm0.12$ &$437.99\pm24.64$\phn&$235.16\pm13.22$\phn& \\
&8--14  &\multirow{2}{*}{$\Big\}$$2.47\pm0.28$$^{\ast}$\hspace{3 pt}}&\multirow{2}{*}{$\Big\}$$0.25\pm0.14$$^{\ast}$\hspace{3 pt}}&$122.71\pm13.69$\phn&$63.28\pm7.06$\phn\phn& \\
&14--20 &&&$49.82\pm10.33$\phn\phn&$17.97\pm3.73$\phn\phn& \\
&20--25  &\multirow{2}{*}{$\Big\}$$1.30\pm0.32$$^{\ast}$\hspace{3 pt}}&\multirow{2}{*}
{$\Big\}$$0.25\pm0.20$$^{\ast}$\hspace{3 pt}}&$4.68\pm4.18$\phn\phn&$3.26\pm2.91$\phn\phn& \\
&25--30  &&&$13.14\pm9.58$\phn\phn&$2.88\pm2.09$\phn\phn& \\
\hline
\tablecomments{
 a) Normalization of {\tt apec} thermal model is the integrated line-of-sight 
emission measure, $(1/4\pi)\int n_e n_H ds$, in units of $10^{14}$ cm$^{-5}$ sr$^{-1}$. 
b) Surface brightness of group emission ($0.5-2.0$ keV) in units of 
$10^{-9}$ erg s$^{-1}$ cm$^{-2}$ sr$^{-1}$;
uncertainties in surface brightness are proportional to those of their associated normalizations.
Fe abundances should be multiplied by 0.67 if they are to be compared with other results
using Anders \& Grevesse (1989) solar abundances. \\
$^\ast$ To ensure the stability of our deprojection analysis, we needed to tie the 
temperatures and abundances of regions 3 and 4,
as well as the temperatures and abundances of regions 5 and 6.}
\end{tabular}
}
\end{table}

\begin{table}
\caption{Summary of Systematic Uncertainties}
\small{
\begin{tabular}{lccccccc}
\hline
{\multirow{2}{*}{Tests}}&{$n_{\rm e}$}&{T}&{Entropy}&{Pressure}&{Mass}&{$f_{\rm gas}$}&$c$\\
&($10^{-5}$ cm$^{-3}$)&{(keV)}&{(keV cm$^2$)}&{($10^{-5}$ keV cm$^{-3}$)}&{(10$^{14}$ M$_{\odot}$)}&{$$}&{$$}\\
\hline
Marginalized$^{a}$ &$5.48^{+1.72}_{-2.63}$&\multirow{2}{*}{$1.13^{+0.36}_{-0.19}$$^{\ast}$}&$866^{+414}_{-265}$&$6.20^{+2.90}_{-2.80}$ &$1.69^{+0.30}_{-0.30}$&$0.11^{+0.02}_{-0.02}$& $6.9\pm4.5$\\

best-fit$^{b}$ &$6.38^{+1.69}_{-1.65}$&$$&$710^{+305}_{-230}$&$7.21^{+2.42}_{-2.44}$ &$1.69^{+0.30}_{-0.26}$&$0.12^{+0.02}_{-0.02}$& $6.4\pm3.2$\\
\hline
CXB $+$20\% &$-2.92$&$-0.20$&$+168$&$-3.99$ &$-0.39$&$+0.02$&$+3.4$ \\
CXB $-$20\% &$+2.30$&$+0.45$&\phn$+99$&$+6.49$ &$+0.27$&$-0.01$&$-2.0$ \\
MW $~+$20\% &$ -0.48$&$+0.09$&\phn$+88$ &$-0.01$&$-0.05$&$-0.001$&$-0.1$\\
MW $~-$20\% &$+1.50$&$-0.09$&\phn$-66$ &$-0.42$&$-0.18$&$+0.01$&$+0.8$\\
LB $~~~+$20\%&$-0.23$&$+0.03$&\phn$+37$ &$-0.08$&$-0.08$&$+0.003$&$+0.5$\\
LB $~~~-$20\%&$-0.06$&$-0.01$&\phn\phn$-2$ &$-0.13$&$-0.12$&$+0.007$&$+0.5$\\
NXB $~~+$5\%&$+0.70$&$+0.02$&\phn$-36$ &$+0.93$&$-0.13$&$+0.01$&$+0.7$\\
NXB $~~-$5\%&$ -0.70$&$-0.03$&\phn$+36$ &$-0.97$&$-0.11$&$+0.005$&$+0.5$\\
\hline
CXB-2 low$^{\ast\ast}$& $-0.98$&$-0.04$&\phn$+53$ &$-1.32$&$-0.14$&$+0.008$&$+1.0$\\
CXB-2 high$^{\ast\ast}$&$-2.90$&$-0.20$&$+116$ &$-0.97$&$-0.40$&$+0.02$&$+3.6$\\
\hline
\tablecomments{ 
Gas properties (electron density, temperature, entropy, pressure) 
in the outmost region, as well as enclosed mass, enclosed gas mass fraction, and dark matter 
concentration of ESO~3060170.\\
a) Marginalized values determined with freely varying background components 
(``deprojected (marginalized)" in Table~3). \\
b) Best-fit values determined using fixed background components 
(``deprojected (best-fit)" in Table~3). 
These values are used as baseline for systematic uncertainty test of the relative importance of each background component.\\
$^{\ast}$ Projected value of temperature was adopted as the temperature of the last bin as described in \S2.2 and \S3.1.\\
$^{\ast\ast}$ When two broken power laws are used to model the CXB, 
two peaks are found in the probability distribution for the
normalization of the component with photon index 1.96 below 1.2 keV;
we examined the effects of adopting each of these marginalized values, in turn, on our results. }
\end{tabular}
}
\end{table}

 \begin{figure} 
\epsscale{0.6}
\plotone{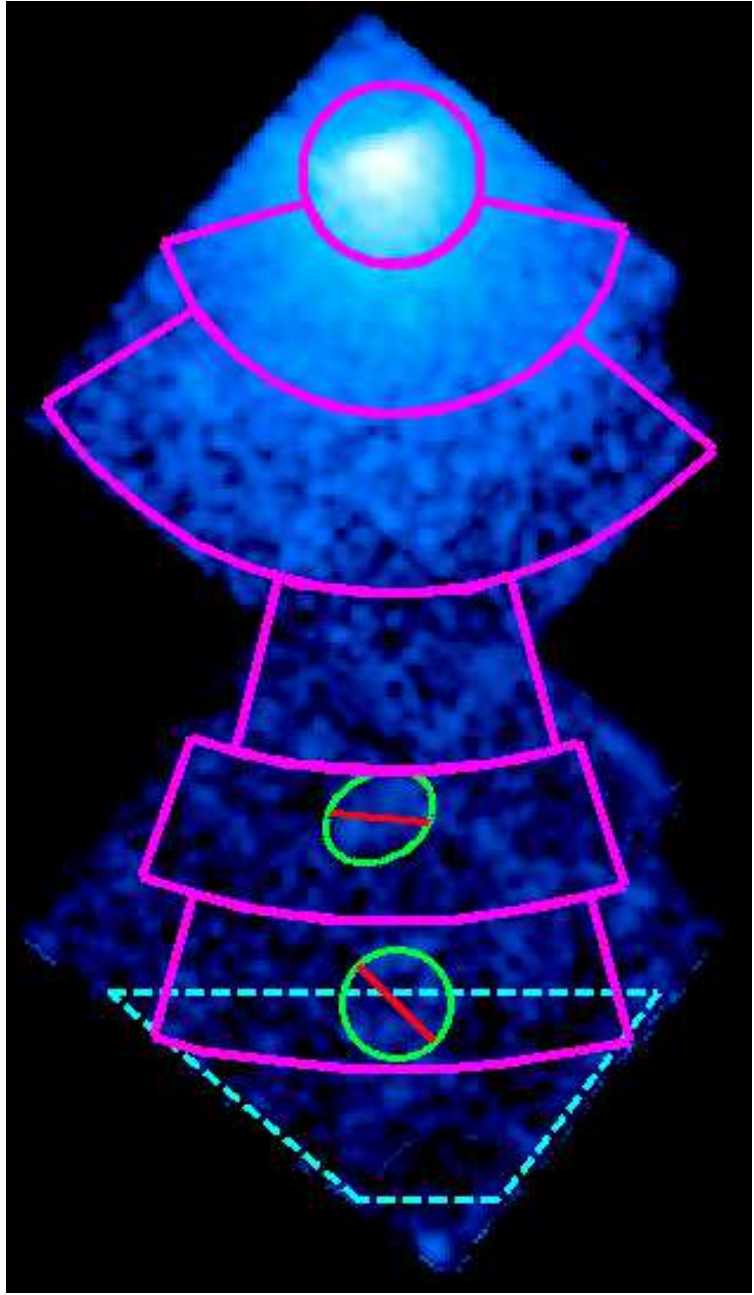}
\figcaption{\label{fig:eso}Mosaic XIS image of ESO~3060170 in 0.5-4.0 keV band with NXB and CXB subtracted. The image is corrected for exposure and vignetting. Pink segments represent regions extracted for spectral analysis. Small green ellipses represent excluded point sources. Cyan trapezoid represents region extracted for ``offset" background spectra. 
[{\sl see the electronic edition of the journal for a color version of this figure.}]}
\end{figure}

\begin{figure}
\vspace{-15mm}
\hspace{5mm}\subfigure{\includegraphics[width=0.5\textwidth]{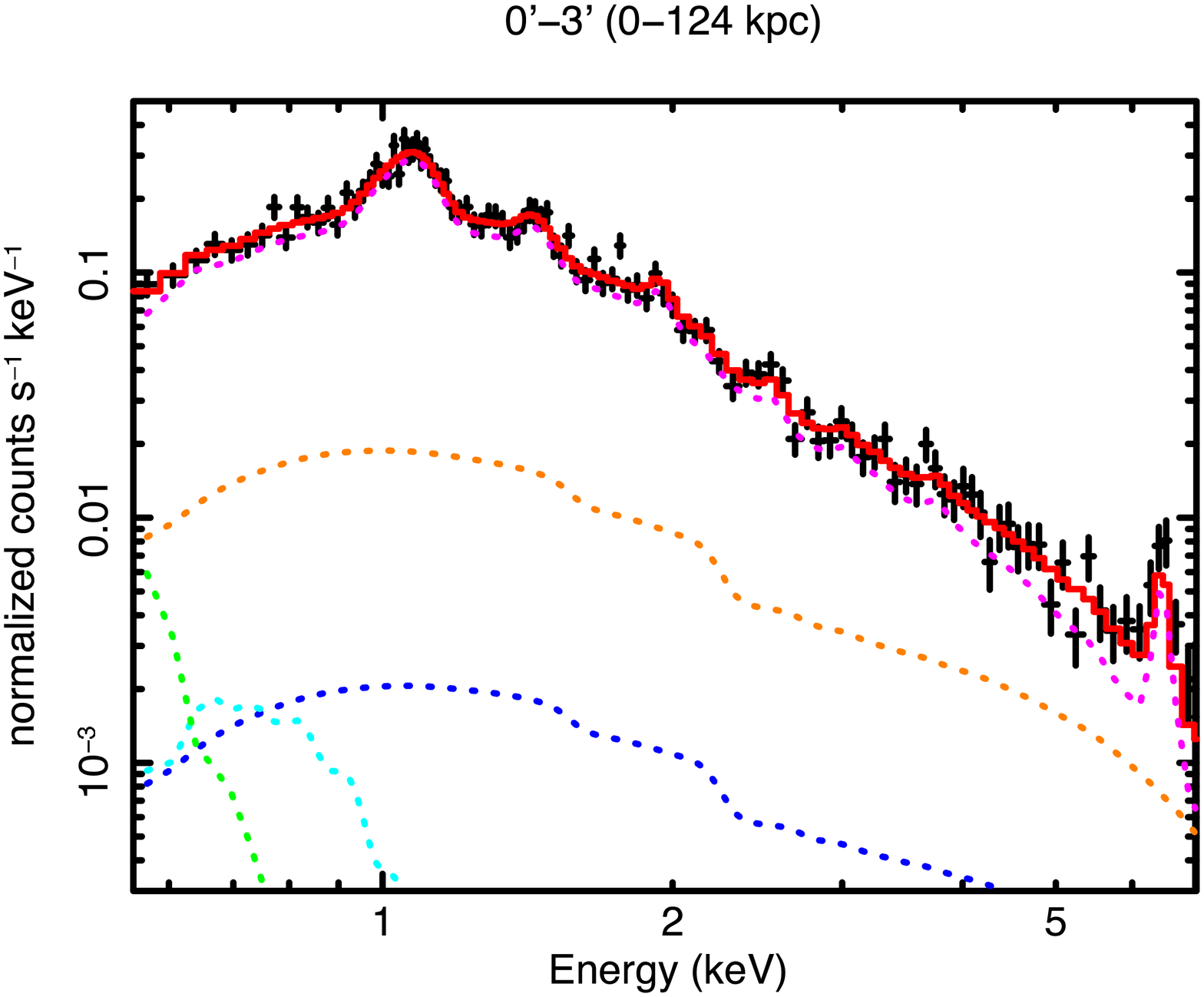}}\hspace{0mm}\subfigure{\includegraphics[width=0.5\textwidth]{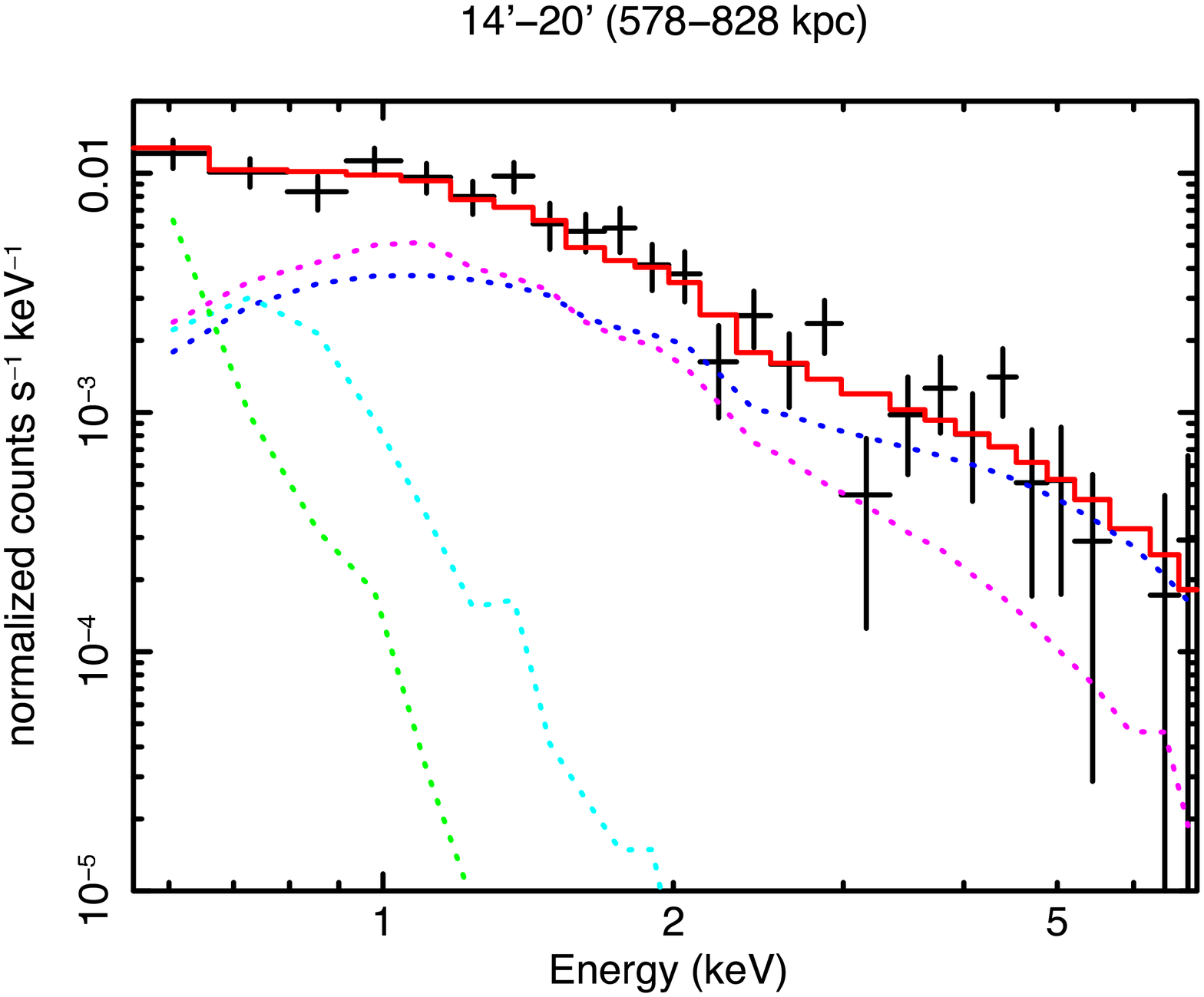}}\\
\vspace{-5mm}
\hspace{5mm}\subfigure{\includegraphics[width=0.5\textwidth]{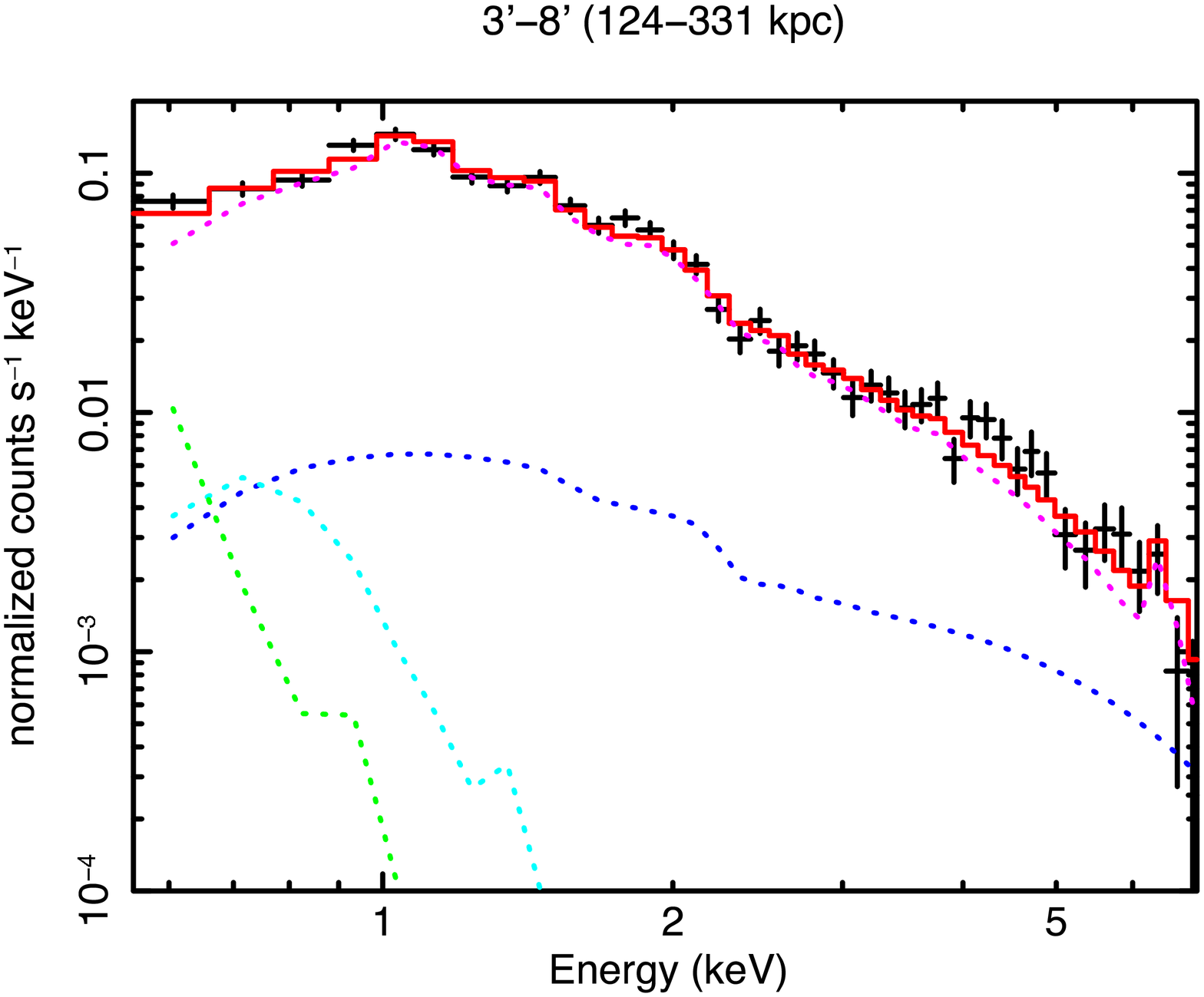}}\hspace{0mm}\subfigure{\includegraphics[width=0.5\textwidth]{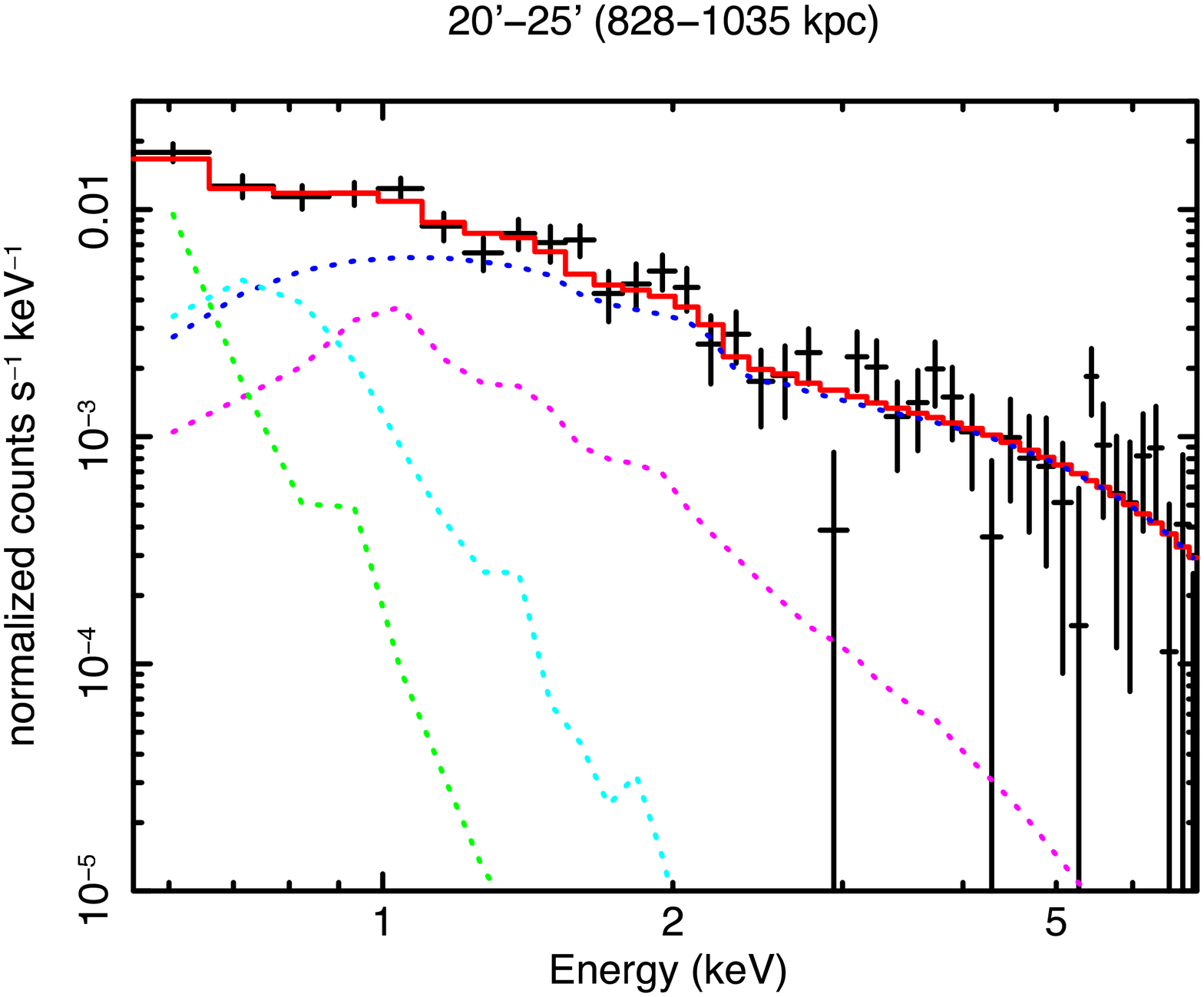}}\\
\vspace{-5mm}
\hspace{5mm}\subfigure{\includegraphics[width=0.5\textwidth]{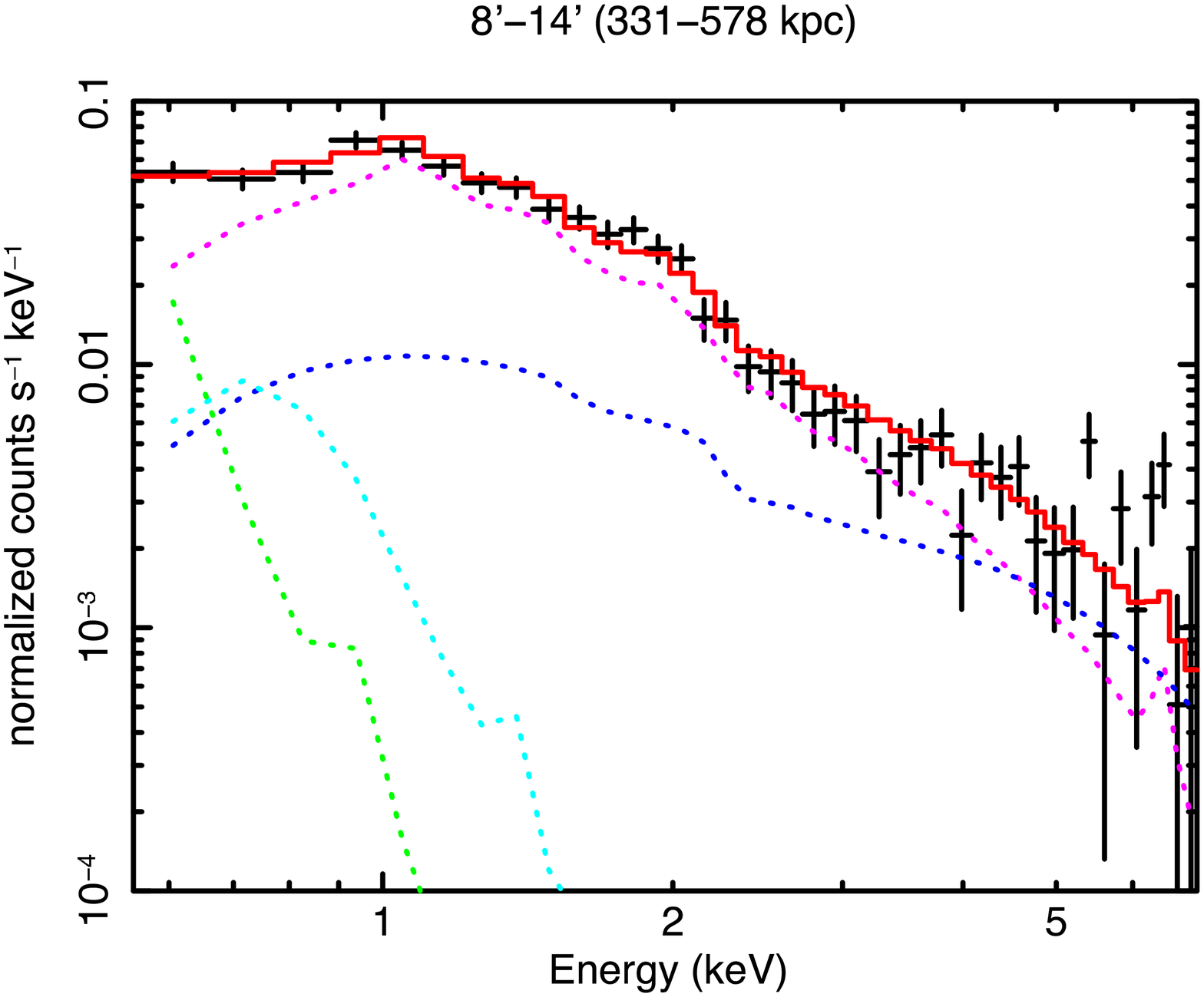}}\hspace{0mm}\subfigure{\includegraphics[width=0.5\textwidth]{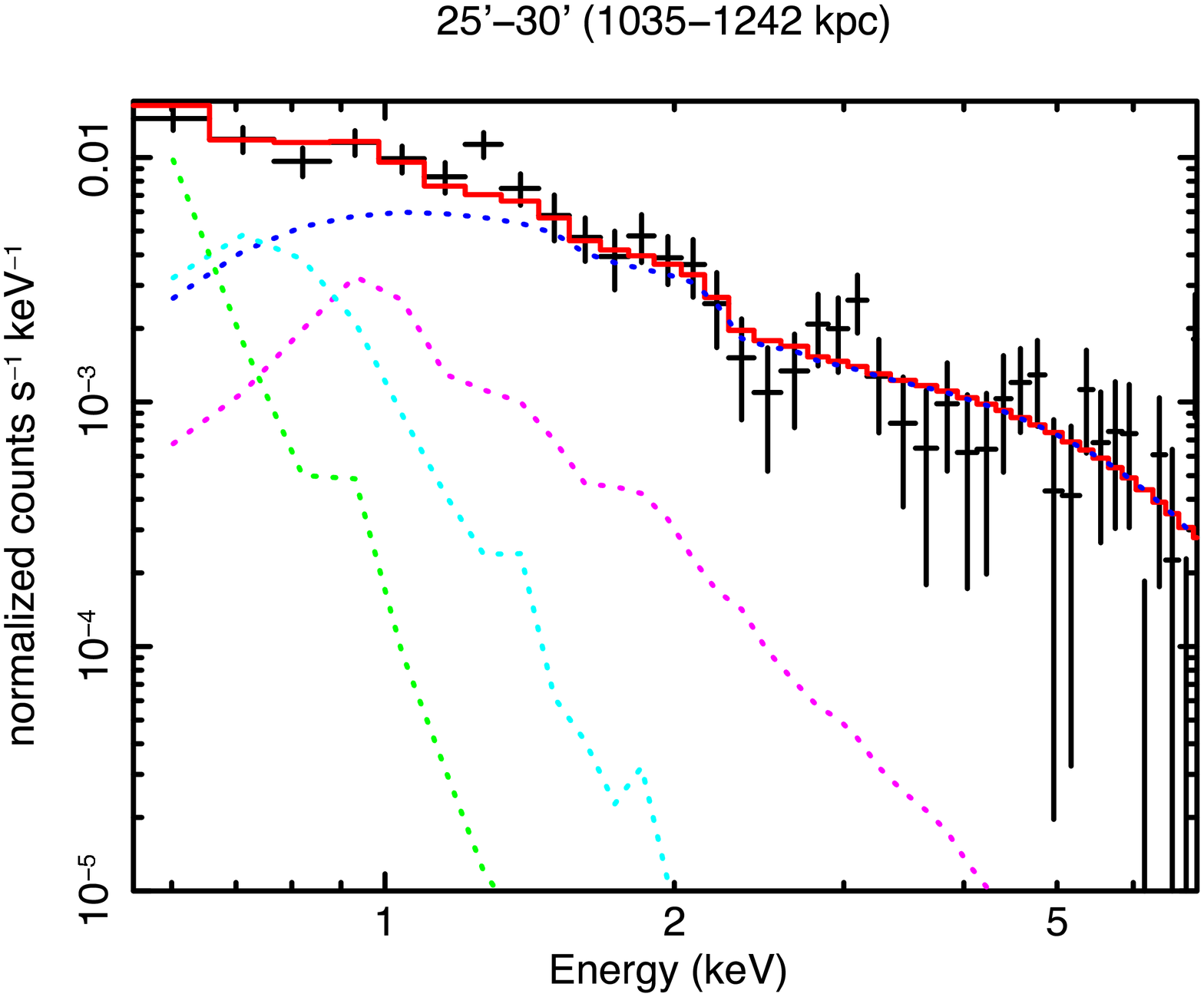}}\\
\figcaption{\label{fig:spectra}{\sl Suzaku} XIS1 spectra for ESO~3060170 from regions 1--6. Instrumental background has been subtracted. 
Black: observed spectra. 
Red: best-fit model. 
Blue: CXB ($\Gamma=1.41$). 
Cyan: MW.  
Green: LB. 
Pink: hot gas emission. 
(Orange: LMXB emission for region 1). 
[{\sl see the electronic edition of the journal for a color version of this figure.}]} 
\end{figure}

\begin{figure}
\epsscale{0.8}
\plotone{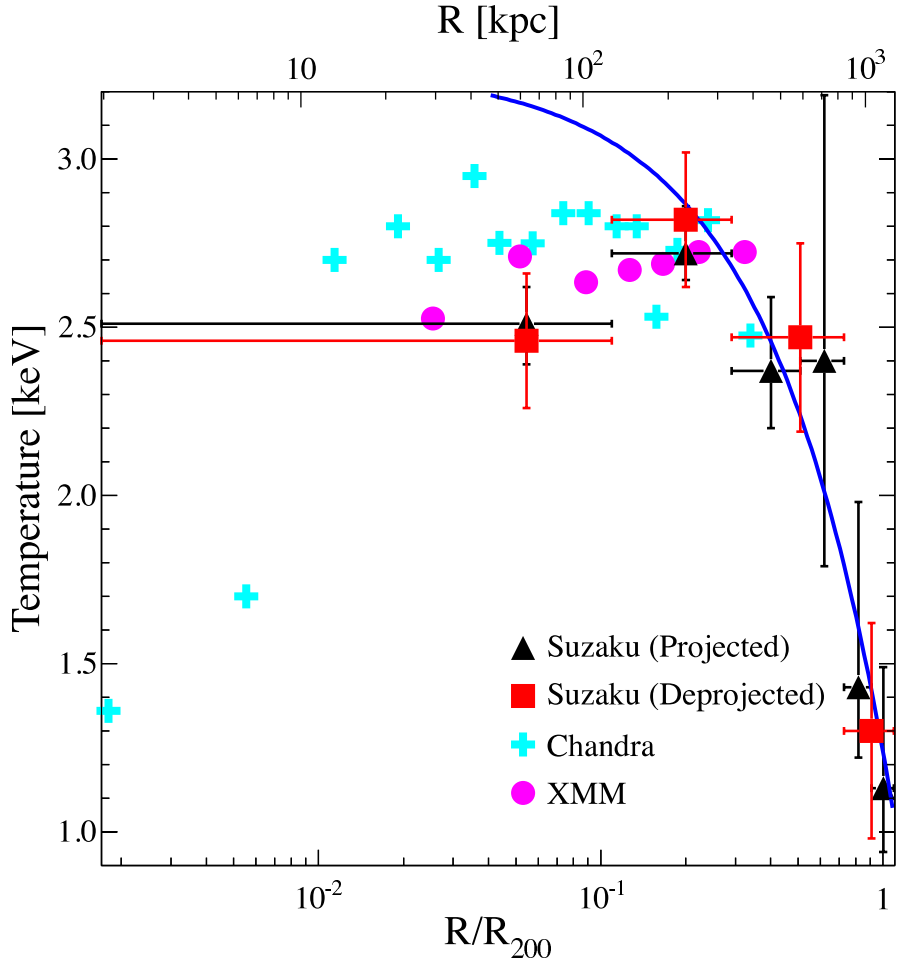}
\figcaption{\label{fig:temperature}Radial temperature profile of ESO 3060170. 
Black triangles: projected temperatures. 
Red squares: deprojected temperatures. 
Blue solid line: the scaled temperature profile by Pratt et al.\ (2007). 
We adopted the deprojected results within 0.8 R$_{200}$ and the projected results beyond 0.8 R$_{200}$ as a global temperature profile. 
Cyan crosses: temperature profile derived with {\sl Chandra} observations (Sun et al.\ 2004). 
Pink circles: temperature profile derived with {\sl XMM-Newton} observations (Sun et al.\ 2004). 
The uncertainties of {\sl Chandra} and {\sl XMM-Newton} measurements are not shown in this plot. [{\sl see the electronic edition of the journal for a color version of this figure.}]}
\end{figure}

\begin{figure}

\plotone{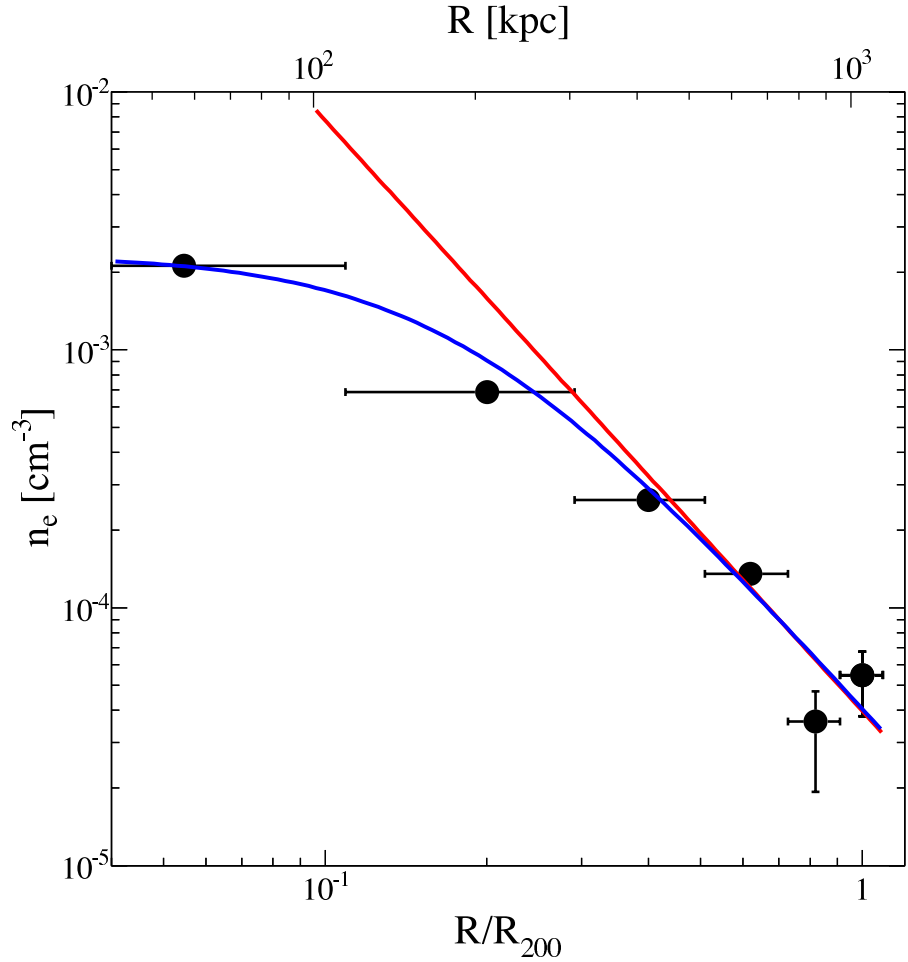}
\figcaption{\label{fig:density}Black dots: deprojected electron density profile of ESO 3060170. Blue line: beta profile fit to deprojected electron density profile, $n_e\propto[1+(r/r_c)^2]^{-3\beta/2}$, where $\beta=0.78, r_c$=188.68 kpc. Red line: power law fit to deprojected electron density profile, $n_e \propto r^{-2.29}$. [{\sl see the electronic edition of the journal for a color version of this figure.}]}
\end{figure}

\clearpage

\begin{figure}
\epsscale{0.8}
\plotone{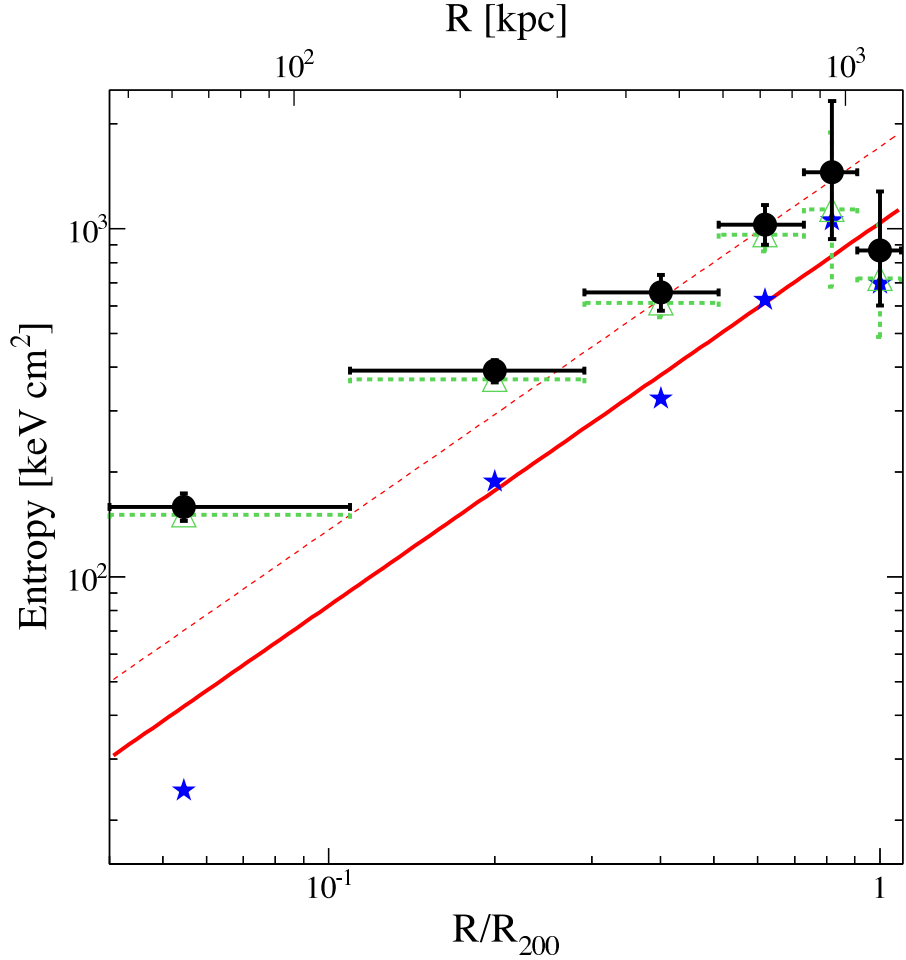}
\figcaption{\label{fig:entropy}Black circles: radial profile of entropy 
($K=n_e^{-2/3}kT$) in ESO 3060170. 
Blue stars: entropy profile corrected the variation of 
$f_{\rm gas}(r)$ with respect to 0.15 (Pratt et al.\ 2010). 
Green triangles: radial profile of entropy determined when background components were fixed in the fitting. 
Red solid line: entropy profile from simulations, 
$K\propto r^{1.1}$; normalization is derived from Voit et al.\ (2005) ($K_{\rm sim}$). 
Red dashed line: entropy profile from simulations, 
$S\propto r^{1.1}$ with normalization determined by fit to intermediate radii ($K_{\rm fit}$). 
[{\sl see the electronic edition of the journal for a color version of this figure.}]}
\end{figure}

\begin{figure} 
\plotone{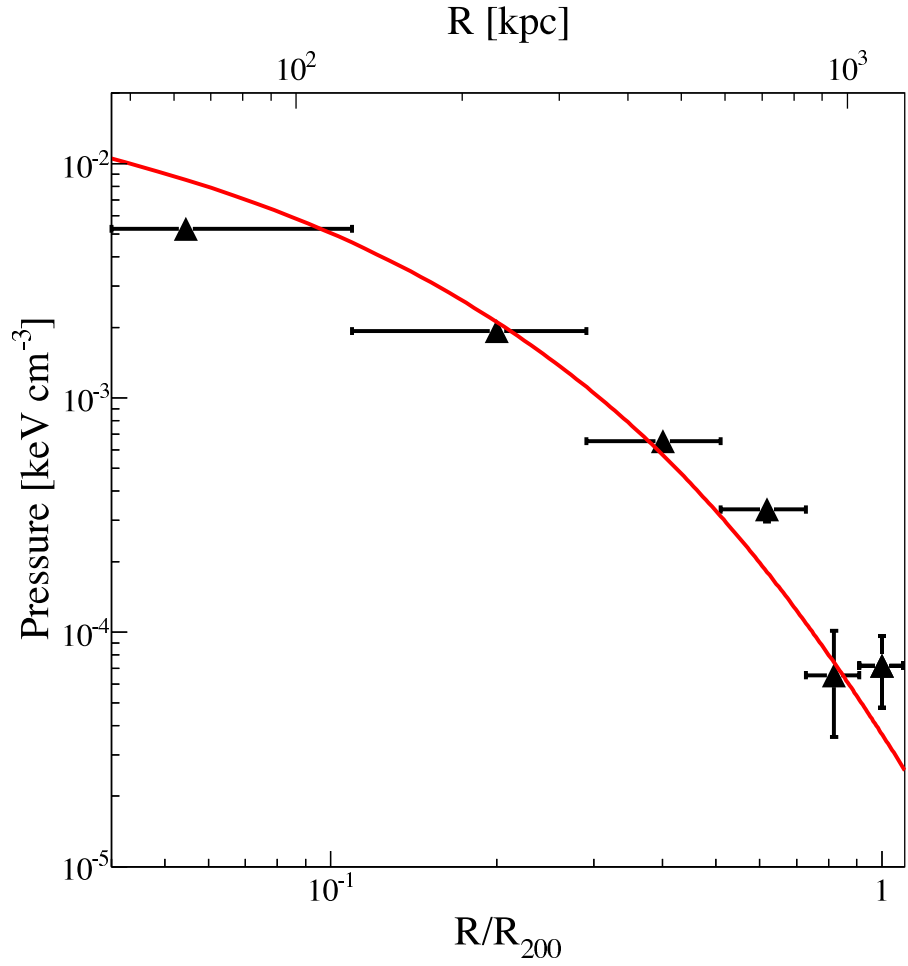}
\figcaption{\label{fig:pressure}Black triangles: radial pressure profile of ESO 3060170. 
Red line: pressure profile derived by Arnaud et al.\ (2010). 
[{\sl see the electronic edition of the journal for a color version of this figure.}]}
\end{figure}

\begin{figure}
\plotone{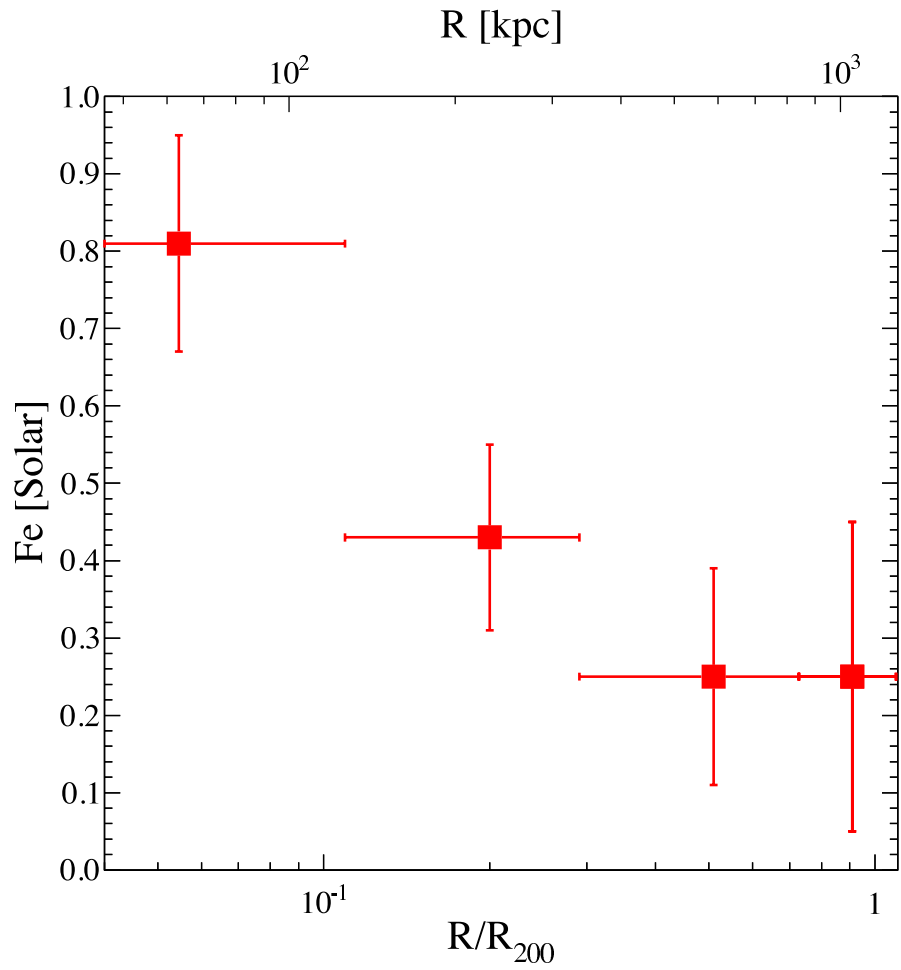}
\figcaption{\label{fig:abundance}Deprojected radial profile of iron abundance in ESO 3060170. }

\end{figure}

\begin{figure} 
\plotone{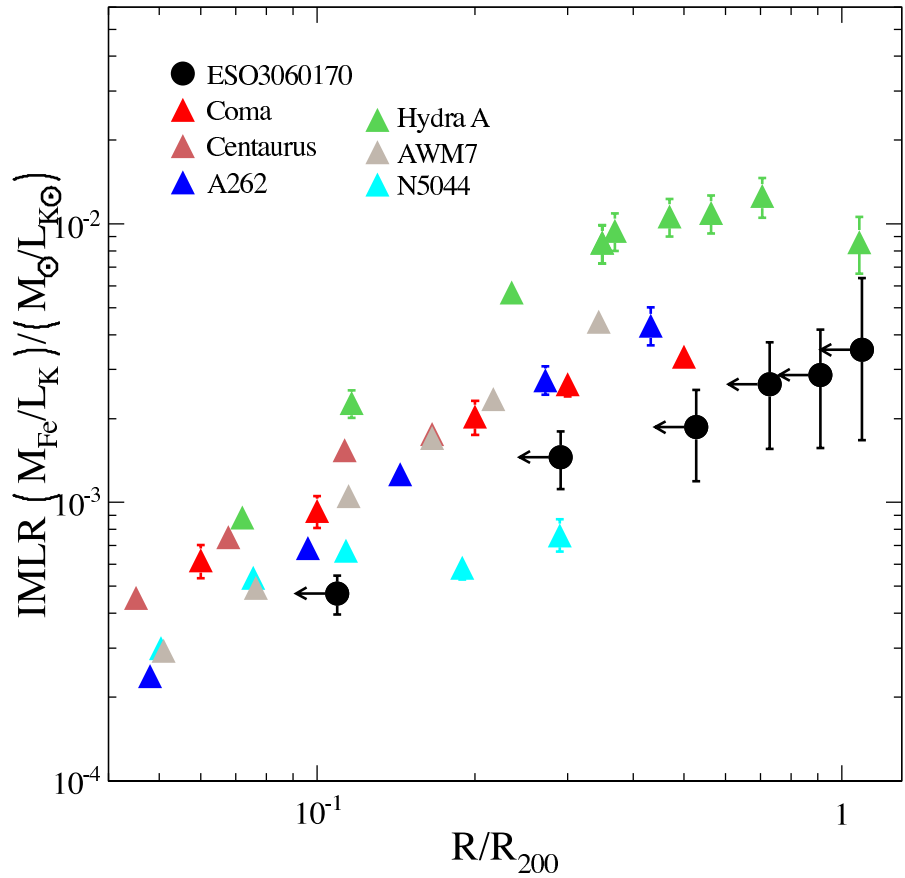}
\figcaption{\label{fig:IMLR}Profile of enclosed iron mass-to-light ratio (IMLR) in ESO 3060170 (black circles), compared to IMLR profiles of other systems (Sato et al.\ 2012); 
the radial coordinates for ESO 3060170 corresponds to the outer endpoints of radial bins. 
Differences in adopted solar abundance tables have been corrected for. 
[{\sl see the electronic edition of the journal for a color version of this figure.}]}
\end{figure}

\begin{figure} 
\plotone{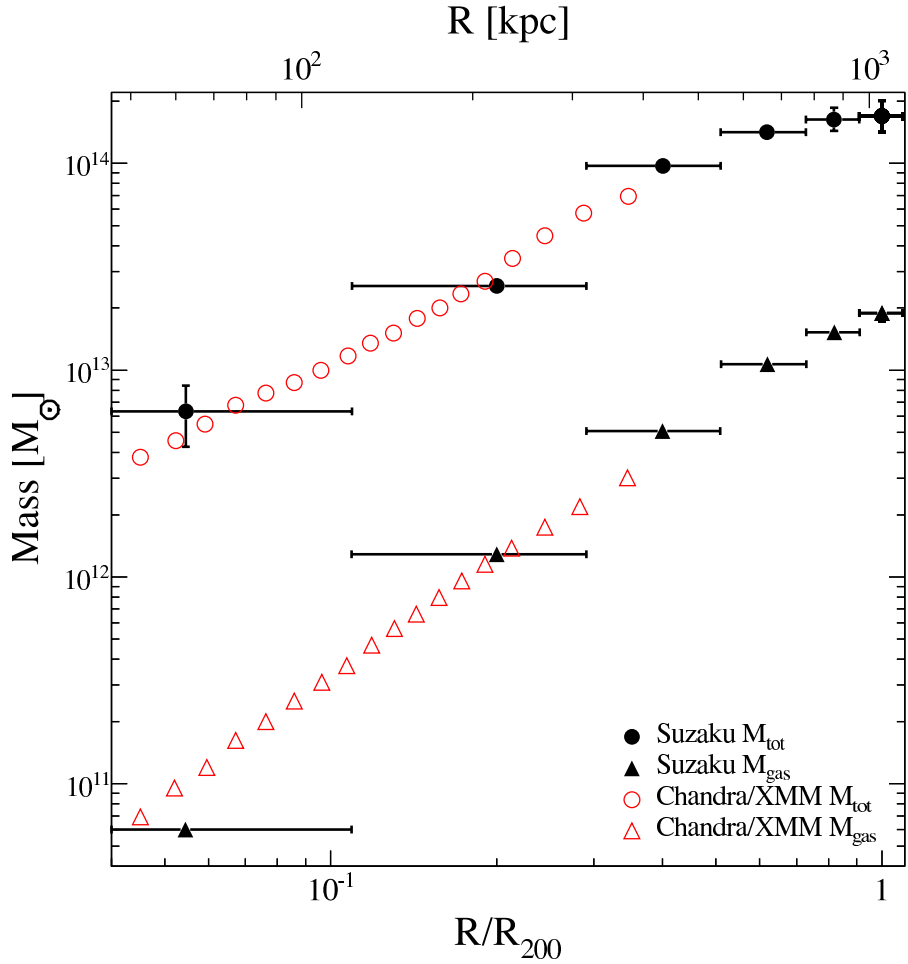}
\figcaption{\label{fig:mass}Black solid triangles: profile of enclosed gas mass in ESO 3060170, as derived from {\sl Suzaku} data. 
Red open triangles: profile of enclosed gas mass in ESO 3060170, as derived from 
{\sl Chandra/XMM} (Sun et al.\ 2004). 
Black solid circles: profile of enclosed total mass, as derived from {\sl Suzaku} data.
Red open circles: profile of enclosed total mass, as derived from 
{\sl Chandra/XMM} data (Sun et al.\ 2004). 
The uncertainties of {\sl Chandra} and {\sl XMM-Newton} measurements are not shown in this plot. 
[{\sl see the electronic edition of the journal for a color version of this figure.}]}
\end{figure}

\begin{figure}
\plotone{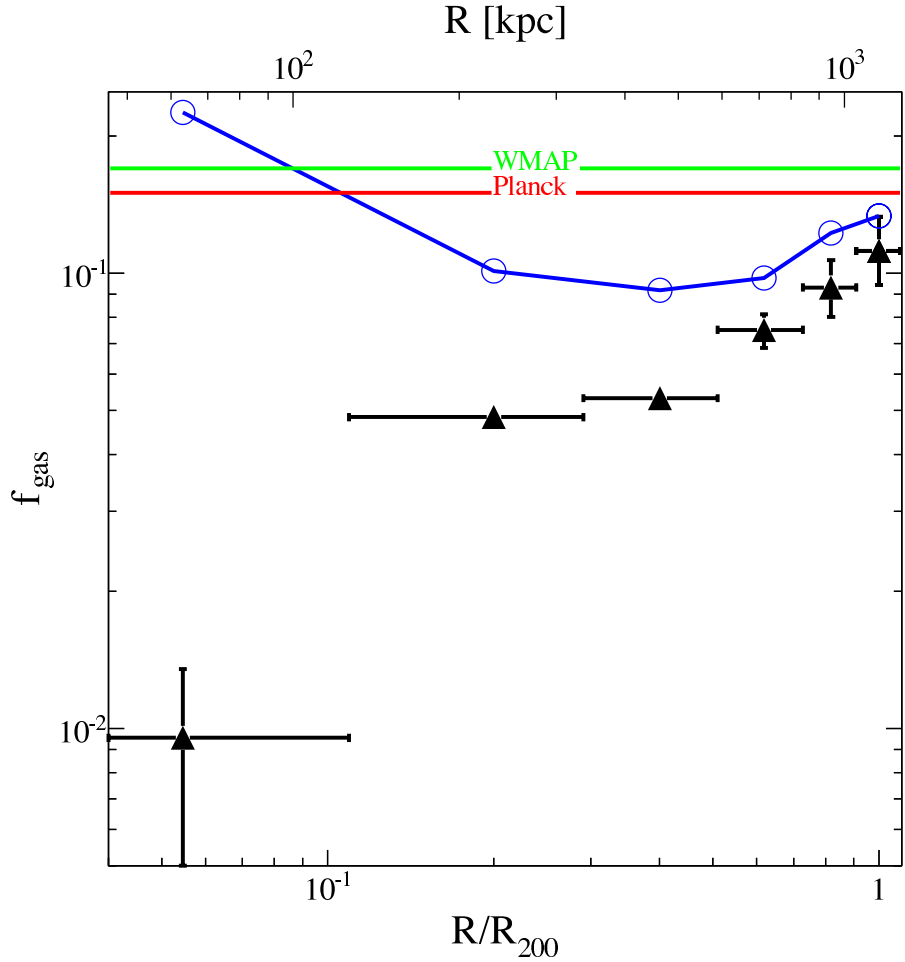}
\figcaption{\label{fig:fg}Black triangles: profile of enclosed gas mass fraction in ESO 3060170. 
Green line: baryon fraction of the universe ($f_b=0.17$), 
as determined by {\sl WMAP} (Hinshaw et al.\ 2009). 
Red line: baryon fraction of the universe ($f_b=0.15$), 
as determined by {\sl Planck} ({\sl Planck} Collaboration, 2013). 
Blue: profile of enclosed baryon fraction (uncertainties are not shown).}
\end{figure}

\begin{figure} 
\plotone{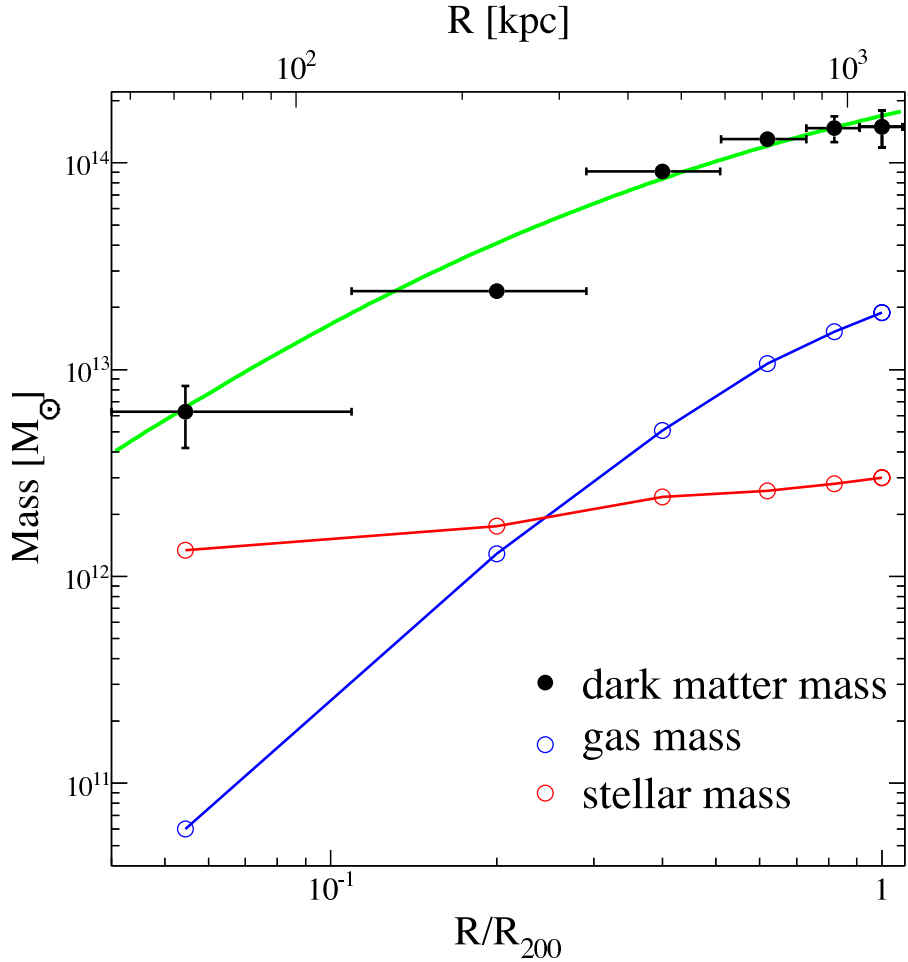}
\figcaption{\label{fig:nfw}Black circles: enclosed dark matter mass profile of ESO 3060170. 
Green line: best-fit NFW dark matter mass profile, with concentration
$c=6.4\pm 4.2$ and scaling radius $r_s=176\pm126$ kpc.
Blue (Red) line: enclosed gas (stellar) mass profile (uncertainties are not shown).  
[{\sl see the electronic edition of the journal for a color version of this figure.}]}
\end{figure}

 \begin{figure}

\hspace{-0mm}
 \epsscale{1.15}

 \plottwo{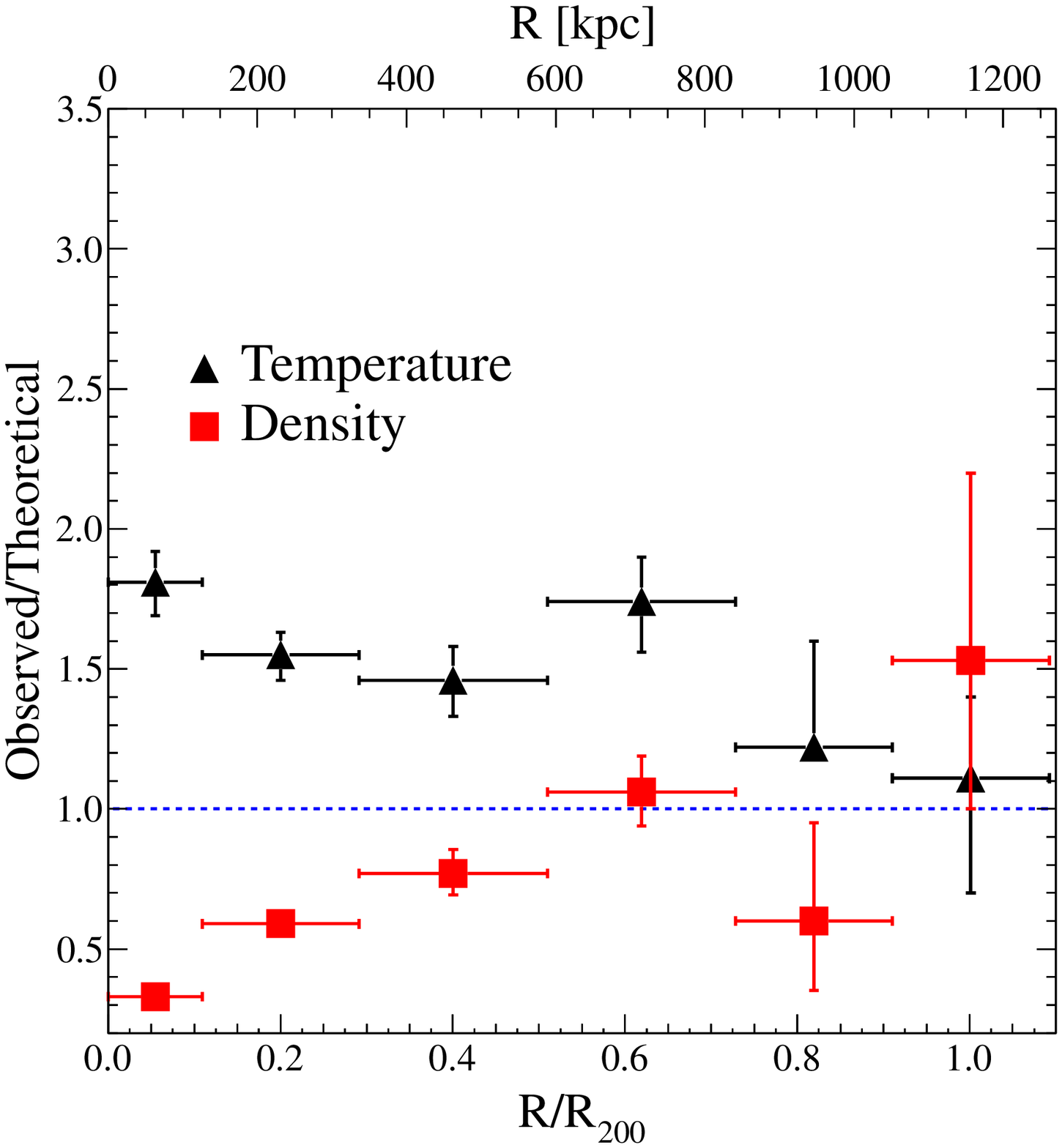}{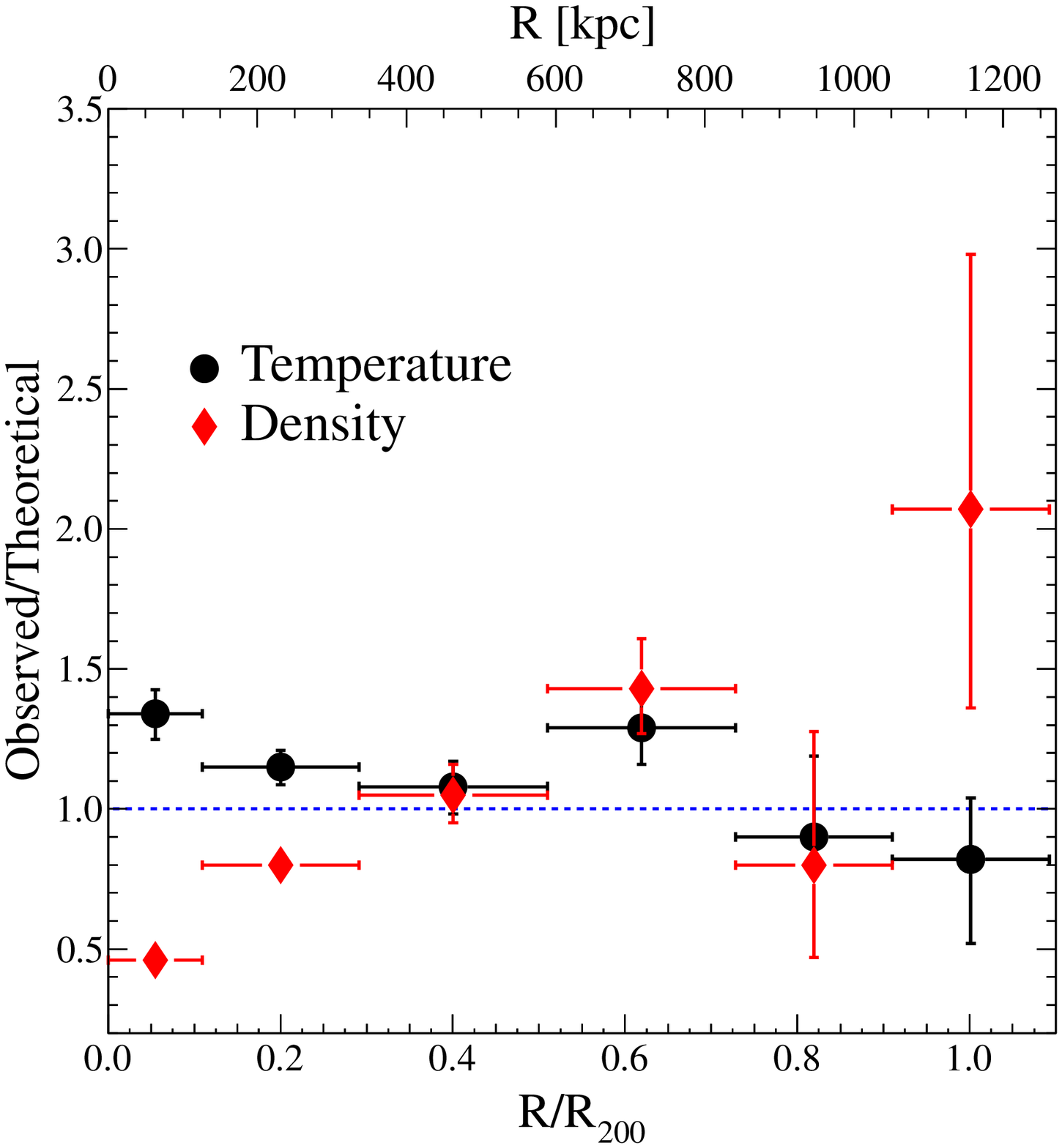}
\put(-450,195){(a) ~$K_{\rm sim}$ baseline:}
\put(-207,195){(b) ~$K_{\rm fit}$ baseline:}
\figcaption{\label{fig:nt}
a) Adopting entropy profile $K_{\rm sim}$ as baseline,
with normalization characterized in Voit et al.\ (2005).
Red squares (black triangles): ratio of observed electron density (temperature) in ESO 3060170 to 
theoretically expected profile.  
Blue dotted line: Observed/Theoretical = 1.
b): Adopting entropy profile $K_{\rm fit}$ as baseline,
with normalization determined by fitting observed profile at intermediate radii.
Red diamonds (black circles): ratio of observed electron density (temperature) to 
theoretically expected profile.  
 [{\sl see the electronic edition of the journal for a color version of this figure.}]}
\end{figure}

\begin{figure}
\epsscale{1.05}
\hspace{-10mm}
\plotone{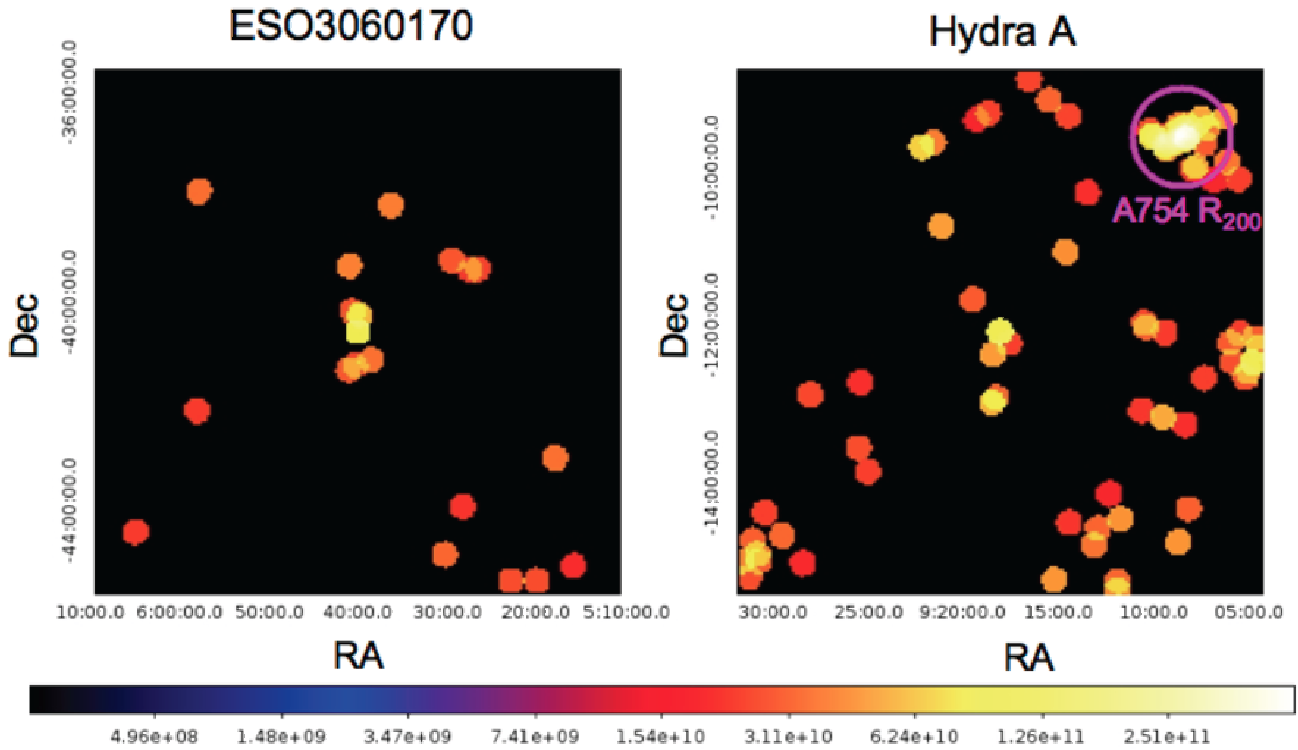}
\figcaption{\label{fig:map}Maps of galaxies around ESO~3060170 ({\sl left}) 
and Hydra A ({\sl right}) on a scale of 25 Mpc, extracted from the 6dF galaxy survey.
Galaxies are color-coded by their $R$-band luminosities. 
Pink circle: $R_{200}$ (= 2.3 Mpc) of Abell~754 (Sivakoff et al.\ 2008).
[{\sl see the electronic edition of the journal for a color version of this figure.}]}
\end{figure}

\clearpage

\begin{figure}
\centering
\vspace{-5.0mm}
\hspace{-0mm}
\subfigure{\includegraphics[width=0.45\textwidth]{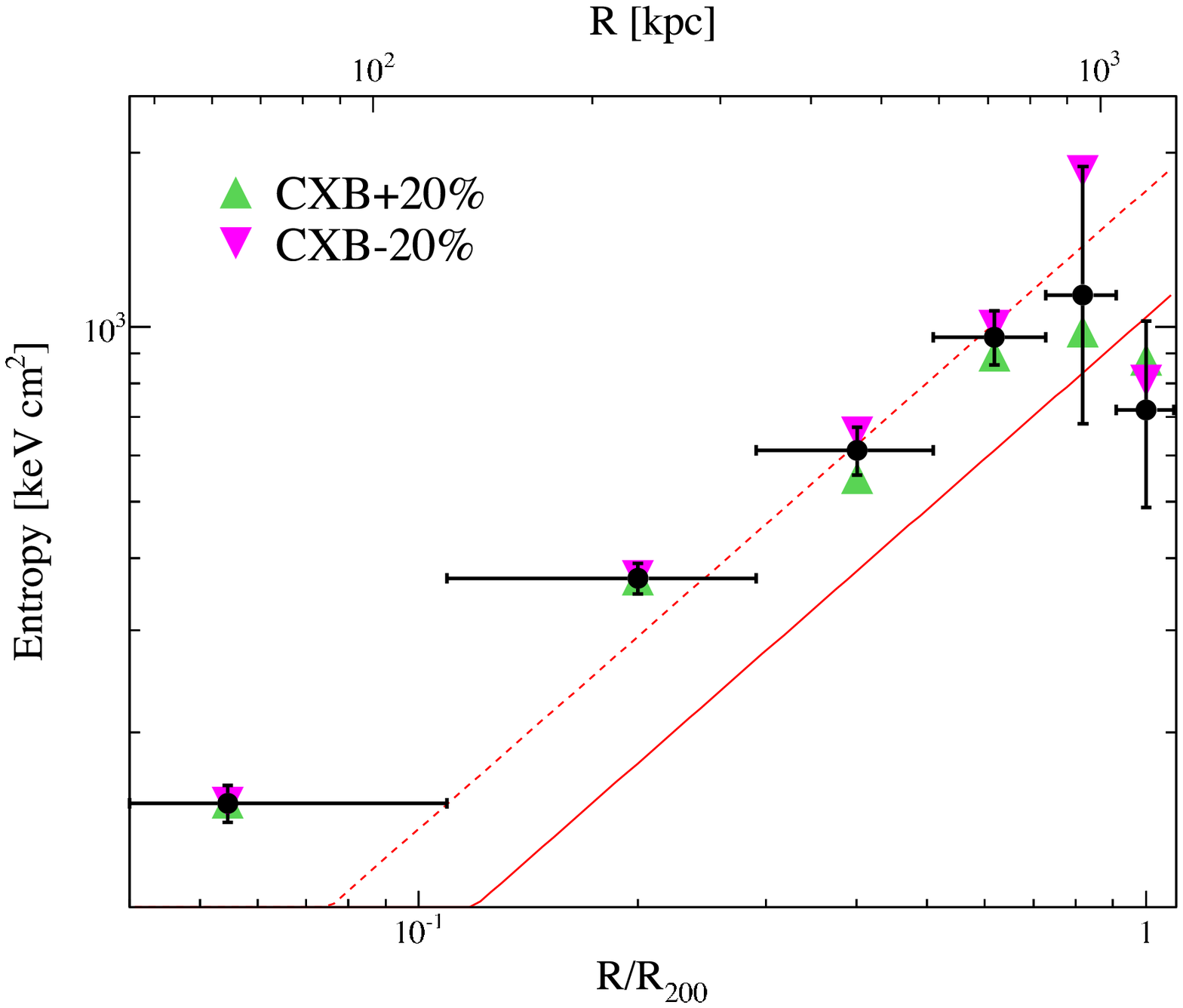}}
\subfigure{\includegraphics[width=0.45\textwidth]{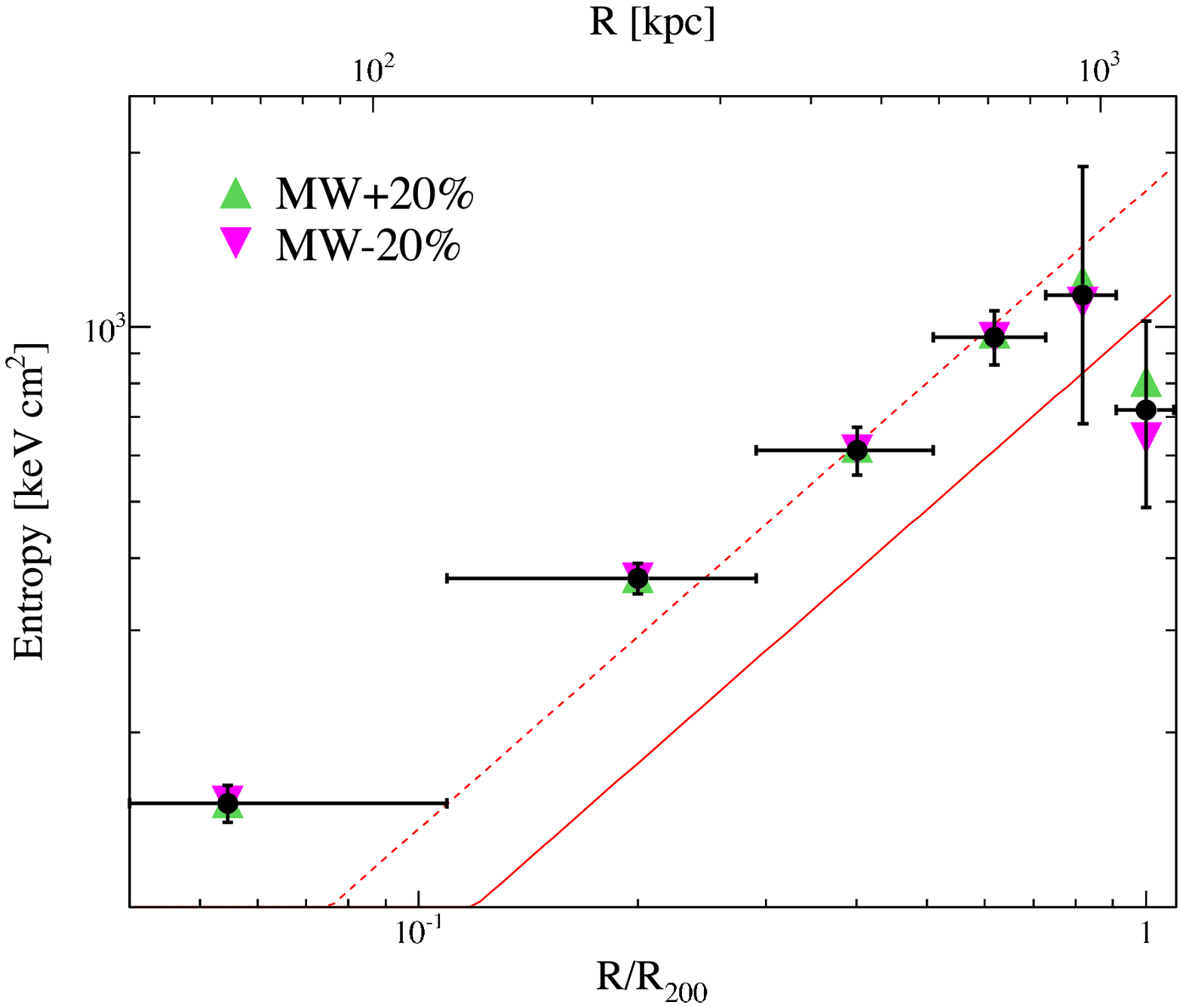}}\\
\hspace{-0mm}
\subfigure{\includegraphics[width=0.45\textwidth]{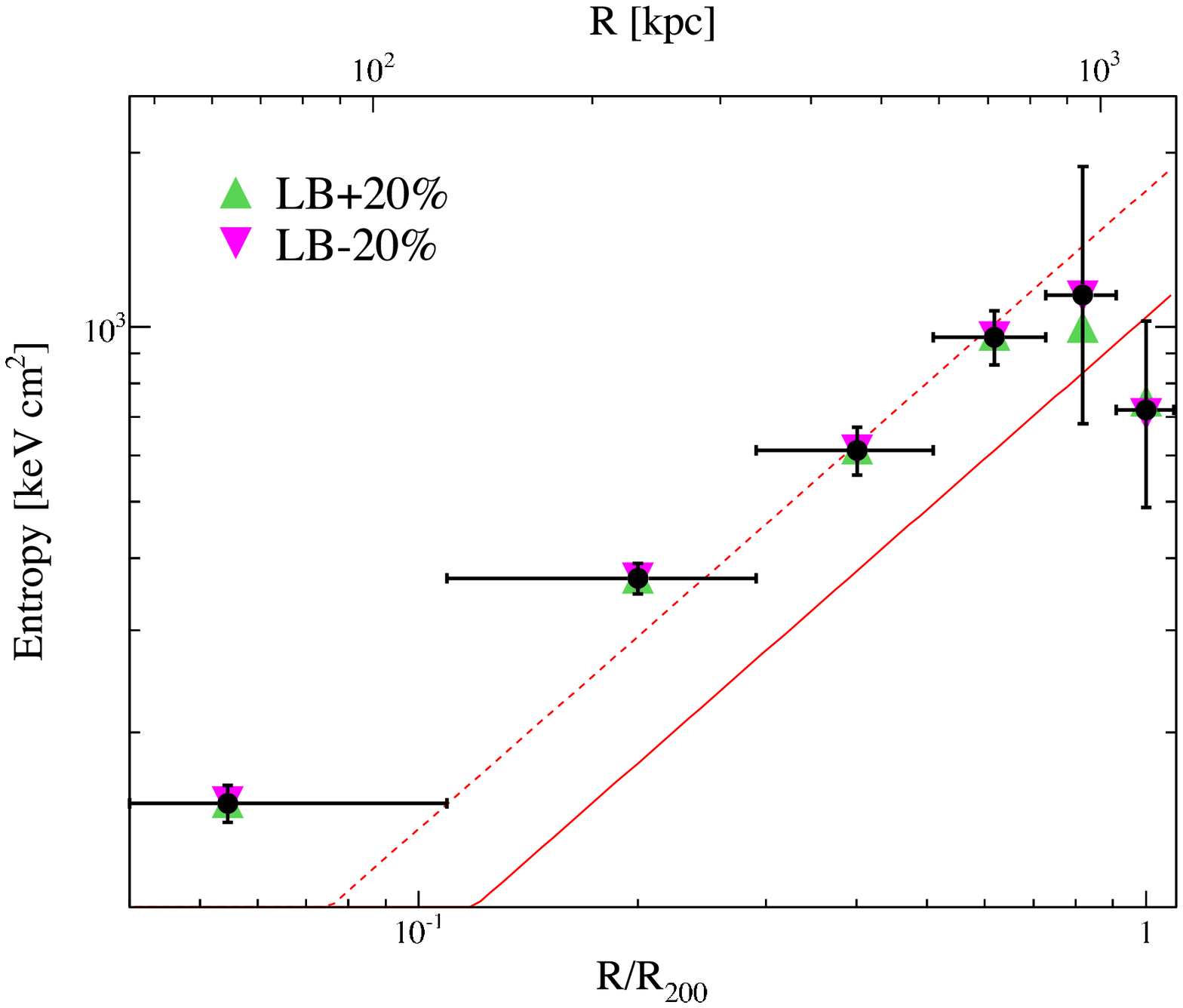}}
\subfigure{\includegraphics[width=0.45\textwidth]{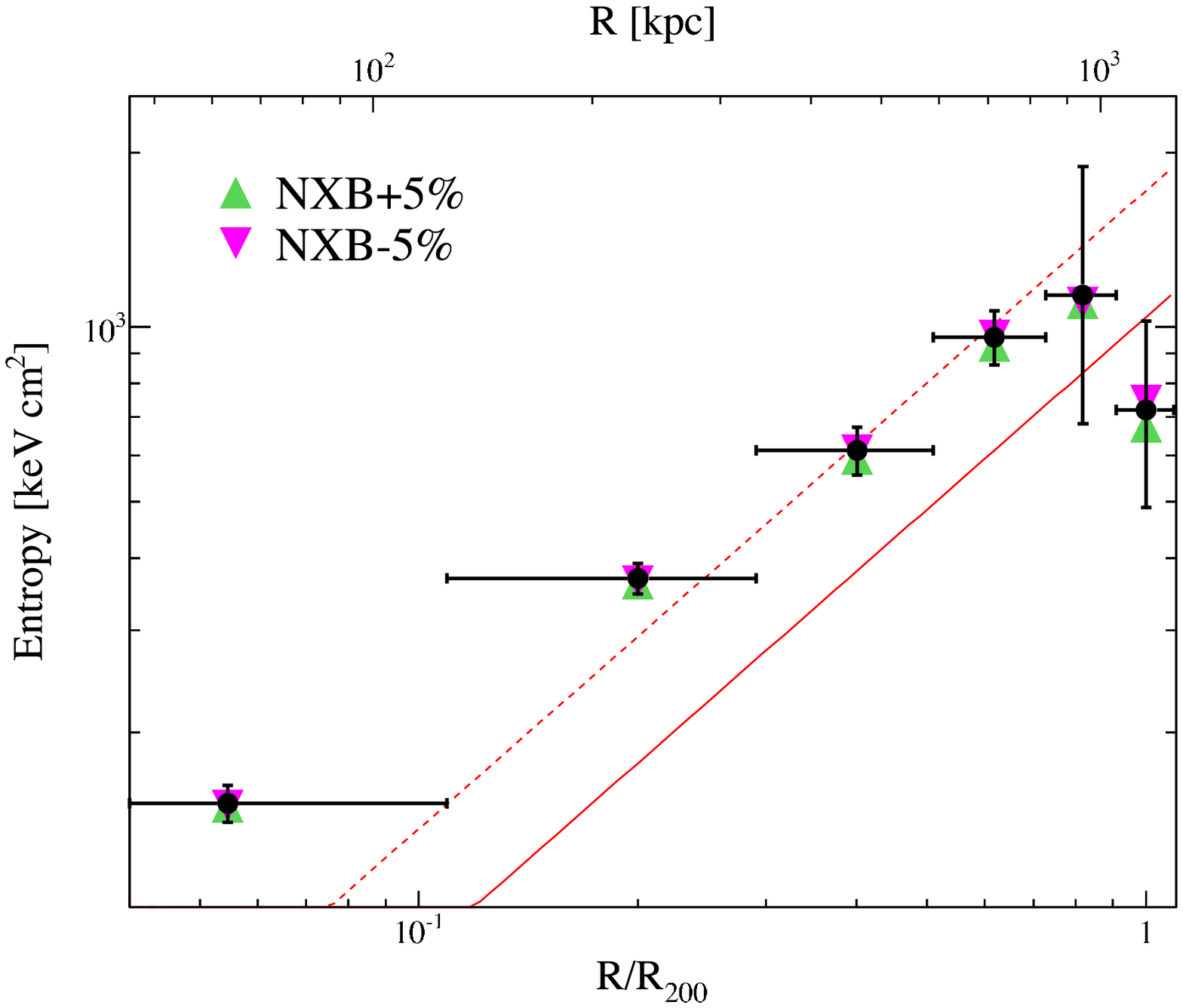}}\\
\hspace{-0mm}
\subfigure{\includegraphics[width=0.45\textwidth]{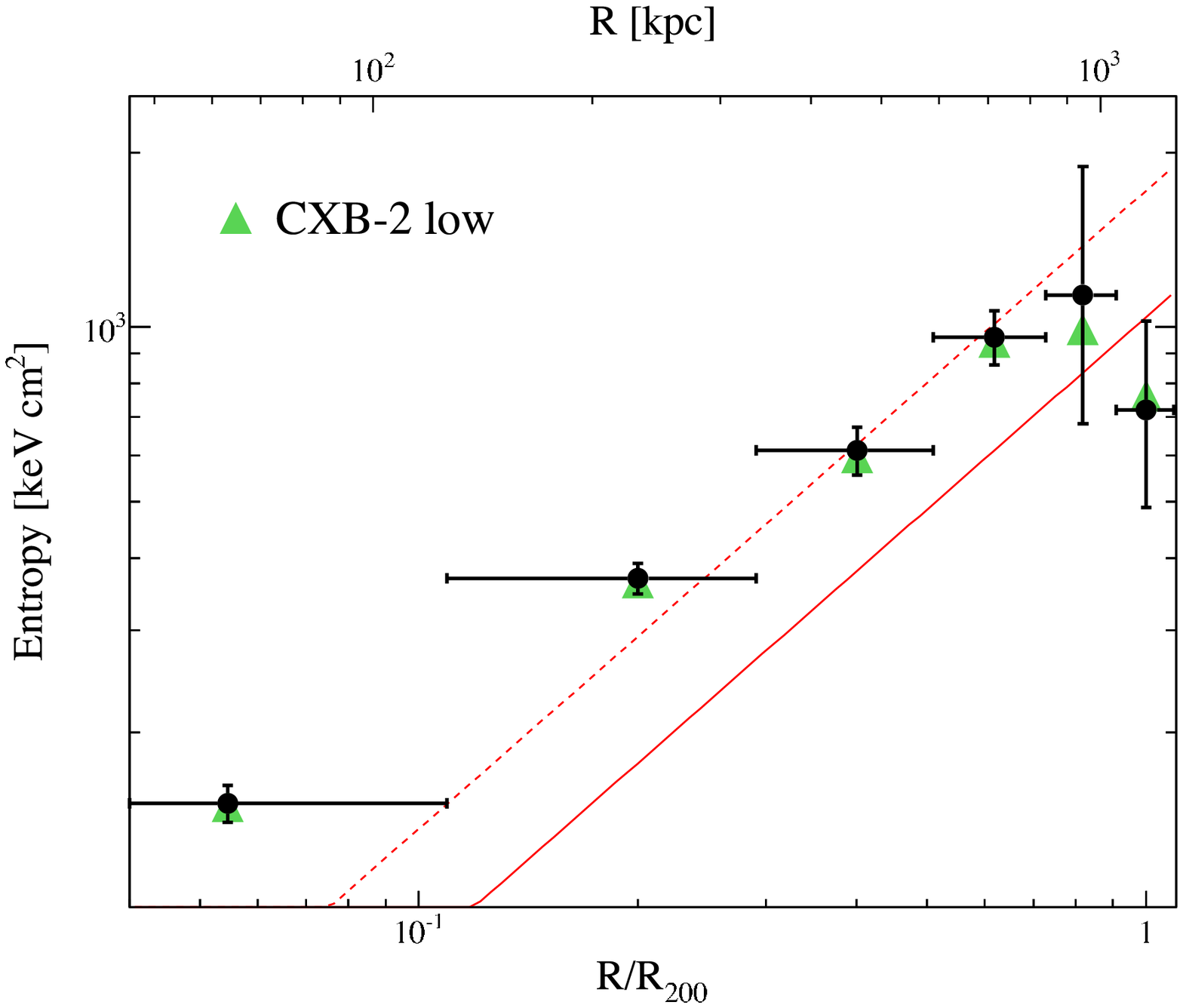}}
\subfigure{\includegraphics[width=0.45\textwidth]{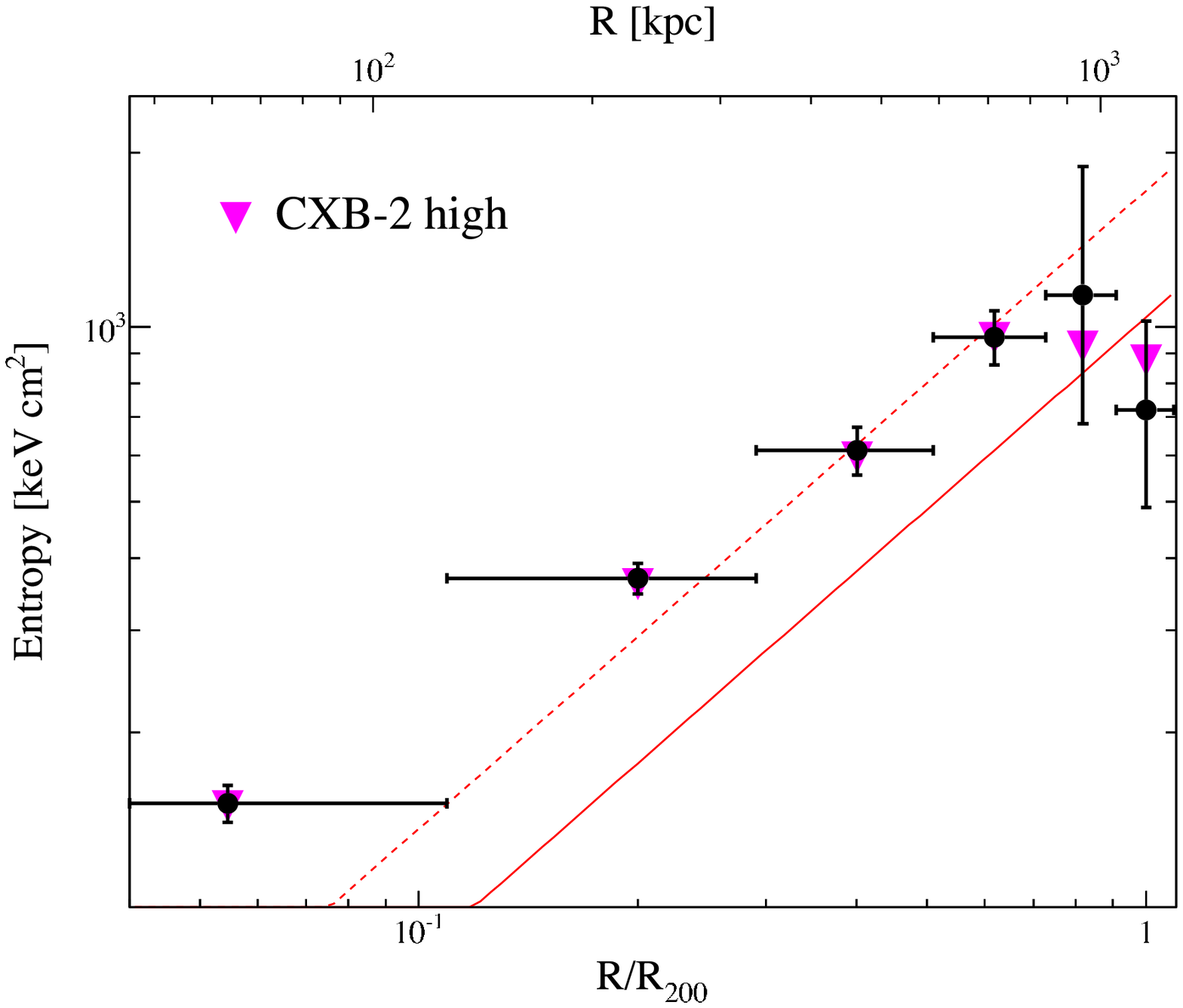}}\\
\figcaption{\label{fig:entropy_sys}Black circles: entropy profile 
($K=n_e^{-2/3}kT$) of ESO 3060170 determined when the background components were fixed in the fitting. 
Red solid (dashed) line: theoretical entropy profile $\propto r^{1.1}$ with normalization 
determined from simulations (fitting). 
{\sl top four figures:}
green (pink) triangles are entropy profiles after increasing (decreasing) 
the levels of the CXB, MW, LB, and NXB background components, respectively.
{\sl bottom two figures:}  
green (pink) triangles are entropy profiles after using two broken power laws 
for CXB modeling and adopting the low (high) marginalized value for the 
normalization of the bright AGN component.
[{\sl see the electronic edition of the journal for a color version of this figure.}]}
\end{figure}

\clearpage

\begin{figure}
\epsscale{0.8}
\plotone{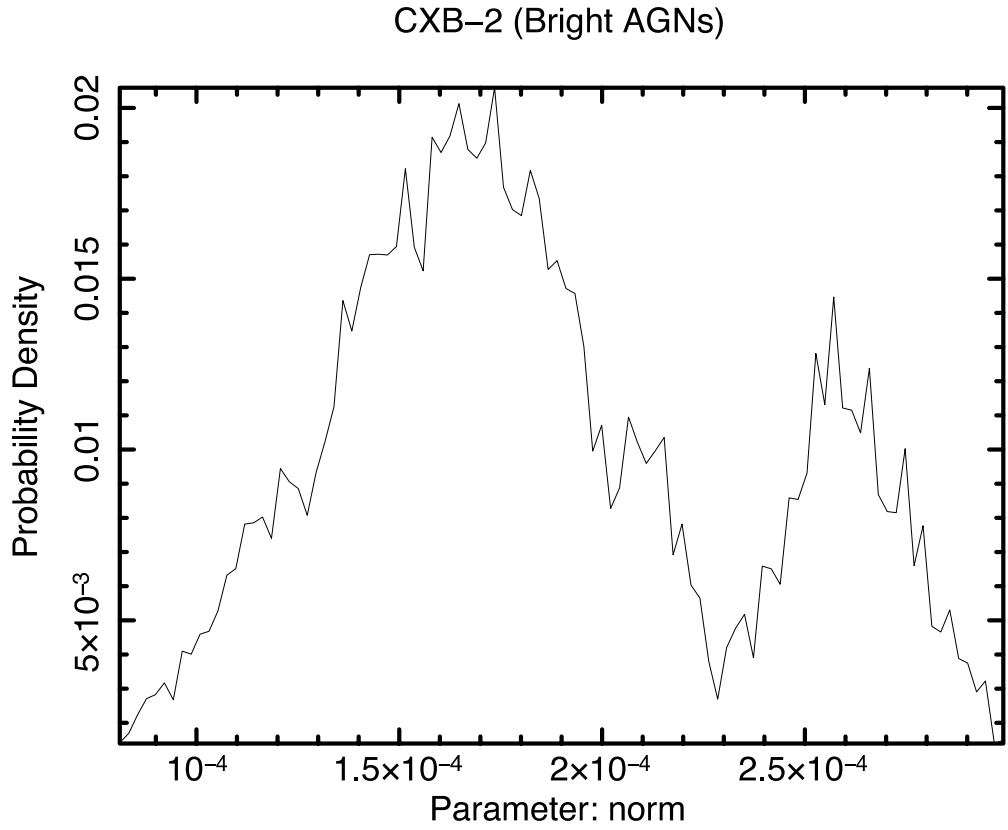}
\figcaption{\label{fig:margin}
The double-peaked probability distribution for the bright AGN component 
($\Gamma_1=1.96$ for $E< 1.2$ keV; $\Gamma_2=1.4$ for $E> 1.2$ keV). 
The normalization of the faint AGN broken power law component is fixed
when two broken power laws are used to respectively model the faint and 
bright AGN components of the CXB.} 
\end{figure}

\end{document}